

%
\expandafter\ifx\csname phyzzx\endcsname\relax
 \message{It is better to use PHYZZX format than to
          \string\input\space PHYZZX}\else
 \wlog{PHYZZX macros are already loaded and are not
          \string\input\space again}%
 \endinput \fi
\catcode`\@=11 
\let\rel@x=\relax
\let\n@expand=\relax
\def\pr@tect{\let\n@expand=\noexpand}
\let\protect=\pr@tect
\let\gl@bal=\global
%
%
%
\newfam\cpfam
\newdimen\b@gheight             \b@gheight=12pt
\newcount\f@ntkey               \f@ntkey=0
\def\f@m{\afterassignment\samef@nt\f@ntkey=}
\def\samef@nt{\fam=\f@ntkey \the\textfont\f@ntkey\rel@x}
\def\setstr@t{\setbox\strutbox=\hbox{\vrule height 0.85\b@gheight
                                depth 0.35\b@gheight width\z@ }}
%
%
%
%
%
\newfam\ssfam   
%
\font\seventeenrm =cmr17
\font\fourteenrm  =cmr10 scaled\magstep2 
\font\twelverm    =cmr12
\font\ninerm      =cmr9
\font\sixrm       =cmr6
%

\font\fourteenss  =cmss10 scaled\magstep2
\font\twelvess    =cmss10 scaled\magstep1
\font\tenss       =cmss10
%

%

%

\font\fourteenbf  =cmbx10 scaled\magstep2 
\font\twelvebf    =cmbx12
\font\ninebf      =cmbx9
\font\sixbf       =cmbx6
%
\font\seventeeni  =cmmi12 scaled\magstep2    \skewchar\seventeeni='177
\font\fourteeni   =cmmi10 scaled\magstep2     \skewchar\fourteeni='177
\font\twelvei     =cmmi12                       \skewchar\twelvei='177
\font\ninei       =cmmi9                          \skewchar\ninei='177
\font\sixi        =cmmi6                           \skewchar\sixi='177
%
\font\seventeensy =cmsy10 scaled\magstep3    \skewchar\seventeensy='60
\font\fourteensy  =cmsy10 scaled\magstep2     \skewchar\fourteensy='60
\font\twelvesy    =cmsy10 scaled\magstep1       \skewchar\twelvesy='60
\font\ninesy      =cmsy9                          \skewchar\ninesy='60
\font\sixsy       =cmsy6                           \skewchar\sixsy='60
%

\font\fourteenex  =cmex10 scaled\magstep2
\font\twelveex    =cmex10 scaled\magstep1
%

\font\fourteensl  =cmsl10 scaled\magstep2
\font\twelvesl    =cmsl12
\font\ninesl      =cmsl9
%

\font\fourteenit  =cmti10 scaled\magstep2 
\font\twelveit    =cmti12
\font\nineit      =cmti9
%
\font\fourteentt  =cmtt10 scaled\magstep2
\font\twelvett    =cmtt12
%
\font\fourteencp  =cmcsc10 scaled\magstep2
\font\twelvecp    =cmcsc10 scaled\magstep1
\font\tencp       =cmcsc10
\def\fourteenf@nts{\relax
    \textfont0=\fourteenrm          \scriptfont0=\tenrm
      \scriptscriptfont0=\sevenrm
    \textfont1=\fourteeni           \scriptfont1=\teni
      \scriptscriptfont1=\seveni
    \textfont2=\fourteensy          \scriptfont2=\tensy
      \scriptscriptfont2=\sevensy
    \textfont3=\fourteenex          \scriptfont3=\twelveex
      \scriptscriptfont3=\tenex
    \textfont\itfam=\fourteenit     \scriptfont\itfam=\tenit
    \textfont\slfam=\fourteensl     \scriptfont\slfam=\tensl
    \textfont\bffam=\fourteenbf     \scriptfont\bffam=\tenbf
      \scriptscriptfont\bffam=\sevenbf
    \textfont\ttfam=\fourteentt
    \textfont\cpfam=\fourteencp
    \textfont\ssfam=\fourteenss     \scriptfont\ssfam=\tenss
        \scriptscriptfont\ssfam=\sevenrm }
\def\twelvef@nts{\relax
    \textfont0=\twelverm          \scriptfont0=\ninerm
      \scriptscriptfont0=\sixrm
    \textfont1=\twelvei           \scriptfont1=\ninei
      \scriptscriptfont1=\sixi
    \textfont2=\twelvesy           \scriptfont2=\ninesy
      \scriptscriptfont2=\sixsy
    \textfont3=\twelveex          \scriptfont3=\tenex
      \scriptscriptfont3=\tenex
    \textfont\itfam=\twelveit     \scriptfont\itfam=\nineit
    \textfont\slfam=\twelvesl     \scriptfont\slfam=\ninesl
    \textfont\bffam=\twelvebf     \scriptfont\bffam=\ninebf
      \scriptscriptfont\bffam=\sixbf
    \textfont\ttfam=\twelvett
    \textfont\cpfam=\twelvecp
    \textfont\ssfam=\twelvess    \scriptfont\ssfam=\tenss
        \scriptscriptfont\ssfam=\sixrm }
\def\tenf@nts{\relax
    \textfont0=\tenrm          \scriptfont0=\sevenrm
      \scriptscriptfont0=\fiverm
    \textfont1=\teni           \scriptfont1=\seveni
      \scriptscriptfont1=\fivei
    \textfont2=\tensy          \scriptfont2=\sevensy
      \scriptscriptfont2=\fivesy
    \textfont3=\tenex          \scriptfont3=\tenex
      \scriptscriptfont3=\tenex
    \textfont\itfam=\tenit     \scriptfont\itfam=\seveni  
    \textfont\slfam=\tensl     \scriptfont\slfam=\sevenrm 
    \textfont\bffam=\tenbf     \scriptfont\bffam=\sevenbf
      \scriptscriptfont\bffam=\fivebf
    \textfont\ttfam=\tentt
    \textfont\cpfam=\tencp
    \textfont\ssfam=\tenss \scriptfont\ssfam=\tenss
            \scriptscriptfont\ssfam=\fiverm }
\def\ss{\n@expand\f@m\ssfam}
%
%
%
\def\rm{\n@expand\f@m0 }
\def\mit{\n@expand\f@m1 }         
\def\cal{\n@expand\f@m2 }
\def\it{\n@expand\f@m\itfam}
\def\sl{\n@expand\f@m\slfam}
\def\bf{\n@expand\f@m\bffam}
\def\tt{\n@expand\f@m\ttfam}
\def\caps{\n@expand\f@m\cpfam}    
\def\em@{\rel@x\ifnum\f@ntkey=0 \it \else
        \ifnum\f@ntkey=\bffam \it \else \rm \fi \fi }
\def\em{\n@expand\em@}
\def\fourteenpoint{\fourteenf@nts \samef@nt \b@gheight=14pt \setstr@t }
\def\twelvepoint{\twelvef@nts \samef@nt \b@gheight=12pt \setstr@t }
\def\tenpoint{\tenf@nts \samef@nt \b@gheight=10pt \setstr@t }
\normalbaselineskip = 20pt plus 0.2pt minus 0.1pt
\normallineskip = 1.5pt plus 0.1pt minus 0.1pt
\normallineskiplimit = 1.5pt
\newskip\normaldisplayskip
\normaldisplayskip = 20pt plus 5pt minus 10pt
\newskip\normaldispshortskip
\normaldispshortskip = 6pt plus 5pt
\newskip\normalparskip
\normalparskip = 6pt plus 2pt minus 1pt
\newskip\skipregister
\skipregister = 5pt plus 2pt minus 1.5pt
\newif\ifsingl@
\newif\ifdoubl@
\newif\iftwelv@  \twelv@true
\def\singlespace{\singl@true\doubl@false\spaces@t}
\def\doublespace{\singl@false\doubl@true\spaces@t}
\def\normalspace{\singl@false\doubl@false\spaces@t}
\def\Tenpoint{\tenpoint\twelv@false\spaces@t}
\def\Twelvepoint{\twelvepoint\twelv@true\spaces@t}
\def\spaces@t{\rel@x
      \iftwelv@ \ifsingl@\subspaces@t3:4;\else\subspaces@t1:1;\fi
       \else \ifsingl@\subspaces@t3:5;\else\subspaces@t4:5;\fi \fi
      \ifdoubl@ \multiply\baselineskip by 5
         \divide\baselineskip by 4 \fi }
\def\subspaces@t#1:#2;{\rel@x
      \baselineskip = \normalbaselineskip
      \multiply\baselineskip by #1 \divide\baselineskip by #2
      \lineskip = \normallineskip
      \multiply\lineskip by #1 \divide\lineskip by #2
      \lineskiplimit = \normallineskiplimit
      \multiply\lineskiplimit by #1 \divide\lineskiplimit by #2
      \parskip = \normalparskip
      \multiply\parskip by #1 \divide\parskip by #2
      \abovedisplayskip = \normaldisplayskip
      \multiply\abovedisplayskip by #1 \divide\abovedisplayskip by #2
      \belowdisplayskip = \abovedisplayskip
      \abovedisplayshortskip = \normaldispshortskip
      \multiply\abovedisplayshortskip by #1
        \divide\abovedisplayshortskip by #2
      \belowdisplayshortskip = \abovedisplayshortskip
      \advance\belowdisplayshortskip by \belowdisplayskip
      \divide\belowdisplayshortskip by 2
      \smallskipamount = \skipregister
      \multiply\smallskipamount by #1 \divide\smallskipamount by #2
      \medskipamount = \smallskipamount \multiply\medskipamount by 2
      \bigskipamount = \smallskipamount \multiply\bigskipamount by 4 }
\def\normalbaselines{ \baselineskip=\normalbaselineskip
   \lineskip=\normallineskip \lineskiplimit=\normallineskip
   \iftwelv@\else \multiply\baselineskip by 4 \divide\baselineskip by 5
     \multiply\lineskiplimit by 4 \divide\lineskiplimit by 5
     \multiply\lineskip by 4 \divide\lineskip by 5 \fi }
\Twelvepoint  
\interlinepenalty=50
\interfootnotelinepenalty=5000
\predisplaypenalty=9000
\postdisplaypenalty=500
\hfuzz=1pt
\vfuzz=0.2pt
\newdimen\HOFFSET  \HOFFSET=0pt
\newdimen\VOFFSET  \VOFFSET=0pt
\newdimen\HSWING   \HSWING=0pt
\dimen\footins=8in
%
%
%
\newskip\pagebottomfiller
\pagebottomfiller=\z@ plus \z@ minus \z@
\def\pagecontents{
   \ifvoid\topins\else\unvbox\topins\vskip\skip\topins\fi
   \dimen@ = \dp255 \unvbox255
   \vskip\pagebottomfiller
   \ifvoid\footins\else\vskip\skip\footins\footrule\unvbox\footins\fi
   \ifr@ggedbottom \kern-\dimen@ \vfil \fi }
\def\makeheadline{\vbox to 0pt{ \skip@=\topskip
      \advance\skip@ by -12pt \advance\skip@ by -2\normalbaselineskip
      \vskip\skip@ \line{\vbox to 12pt{}\the\headline} \vss
      }\nointerlineskip}
\def\makefootline{\baselineskip = 1.5\normalbaselineskip
                 \line{\the\footline}}
\newif\iffrontpage
\newif\ifp@genum
\def\nopagenumbers{\p@genumfalse}
\def\pagenumbers{\p@genumtrue}
\pagenumbers
\newtoks\paperheadline
\newtoks\paperfootline
\newtoks\letterheadline
\newtoks\letterfootline
\newtoks\letterinfo
\newtoks\date
\paperheadline={\hfil}
\paperfootline={\hss\iffrontpage\else\ifp@genum\tenrm\folio\hss\fi\fi}
\letterheadline{\iffrontpage \hfil \else
    \rm \ifp@genum page~~\folio\fi \hfil\the\date \fi}
\letterfootline={\iffrontpage\the\letterinfo\else\hfil\fi}
\letterinfo={\hfil}
\def\monthname{\rel@x\ifcase\month 0/\or January\or February\or
   March\or April\or May\or June\or July\or August\or September\or
   October\or November\or December\else\number\month/\fi}
\def\today{\monthname~\number\day, \number\year}
\date={\today}
\headline=\paperheadline 
\footline=\paperfootline 
\countdef\pageno=1      \countdef\pagen@=0
\countdef\pagenumber=1  \pagenumber=1
\def\advancepageno{\gl@bal\advance\pagen@ by 1
   \ifnum\pagenumber<0 \gl@bal\advance\pagenumber by -1
    \else\gl@bal\advance\pagenumber by 1 \fi
    \gl@bal\frontpagefalse  \swing@ }
\def\folio{\ifnum\pagenumber<0 \romannumeral-\pagenumber
           \else \number\pagenumber \fi }
\def\swing@{\ifodd\pagenumber \gl@bal\advance\hoffset by -\HSWING
             \else \gl@bal\advance\hoffset by \HSWING \fi }
\def\footrule{\dimen@=\prevdepth\nointerlineskip
   \vbox to 0pt{\vskip -0.25\baselineskip \hrule width 0.35\hsize \vss}
   \prevdepth=\dimen@ }
\let\footnotespecial=\rel@x
\newdimen\footindent
\footindent=24pt
\def\Textindent#1{\noindent\llap{#1\enspace}\ignorespaces}
\def\Vfootnote#1{\insert\footins\bgroup
   \interlinepenalty=\interfootnotelinepenalty \floatingpenalty=20000
   \singl@true\doubl@false\Tenpoint
   \splittopskip=\ht\strutbox \boxmaxdepth=\dp\strutbox
   \leftskip=\footindent \rightskip=\z@skip
   \parindent=0.5\footindent \parfillskip=0pt plus 1fil
   \spaceskip=\z@skip \xspaceskip=\z@skip \footnotespecial
   \Textindent{#1}\footstrut\futurelet\next\fo@t}

\def\vfootnote#1{\Vfootnote{${#1}$}}
\def\footnote#1{\attach{#1}\vfootnote{#1}}

\let\footsymbol=\star
\newcount\lastf@@t           \lastf@@t=-1
\newcount\footsymbolcount    \footsymbolcount=0
\newif\ifPhysRev
\def\bumpfootsymbolcount{\rel@x
   \iffrontpage \bumpfootsymbolpos \else \advance\lastf@@t by 1
     \ifPhysRev \bumpfootsymbolneg \else \bumpfootsymbolpos \fi \fi
   \gl@bal\lastf@@t=\pagen@ }
\def\bumpfootsymbolpos{\ifnum\footsymbolcount <0
                            \gl@bal\footsymbolcount =0 \fi
    \ifnum\lastf@@t<\pagen@ \gl@bal\footsymbolcount=0
     \else \gl@bal\advance\footsymbolcount by 1 \fi }
\def\bumpfootsymbolneg{\ifnum\footsymbolcount >0
             \gl@bal\footsymbolcount =0 \fi
         \gl@bal\advance\footsymbolcount by -1 }
\def\fd@f#1 {\xdef\footsymbol{\mathchar"#1 }}
\def\generatefootsymbol{\ifcase\footsymbolcount \fd@f 13F \or \fd@f 279
        \or \fd@f 27A \or \fd@f 278 \or \fd@f 27B \else
        \ifnum\footsymbolcount <0 \fd@f{023 \number-\footsymbolcount }
         \else \fd@f 203 {\loop \ifnum\footsymbolcount >5
                \fd@f{203 \footsymbol } \advance\footsymbolcount by -1
                \repeat }\fi \fi }

\def\nonfrenchspacing{\sfcode`\.=3001 \sfcode`\!=3000 \sfcode`\?=3000
        \sfcode`\:=2000 \sfcode`\;=1500 \sfcode`\,=1251 }
\nonfrenchspacing
\newdimen\d@twidth
{\setbox0=\hbox{s.} \gl@bal\d@twidth=\wd0 \setbox0=\hbox{s}
        \gl@bal\advance\d@twidth by -\wd0 }
\def\removehglue{\loop \unskip \ifdim\lastskip >\z@ \repeat }
\def\roll@ver#1{\removehglue \nobreak \count255 =\spacefactor \dimen@=\z@
        \ifnum\count255 =3001 \dimen@=\d@twidth \fi
        \ifnum\count255 =1251 \dimen@=\d@twidth \fi
    \iftwelv@ \kern-\dimen@ \else \kern-0.83\dimen@ \fi
   #1\spacefactor=\count255 }
\def\step@ver#1{\rel@x \ifmmode #1\else \ifhmode
        \roll@ver{${}#1$}\else {\setbox0=\hbox{${}#1$}}\fi\fi }
\def\attach#1{\step@ver{\strut^{\mkern 2mu #1} }}
%
%
%
\newcount\chapternumber      \chapternumber=0
\newcount\sectionnumber      \sectionnumber=0
\newcount\equanumber         \equanumber=0
\let\chapterlabel=\rel@x
\let\sectionlabel=\rel@x
\newtoks\chapterstyle        \chapterstyle={\Number}
\newtoks\sectionstyle        \sectionstyle={\Number}
\newskip\chapterskip         \chapterskip=\bigskipamount
\newskip\sectionskip         \sectionskip=\medskipamount
\newskip\headskip            \headskip=8pt plus 3pt minus 3pt
\newdimen\chapterminspace    \chapterminspace=15pc
\newdimen\sectionminspace    \sectionminspace=10pc
\newdimen\referenceminspace  \referenceminspace=20pc
\newif\ifcn@                 \cn@true
\newif\ifcn@@                \cn@@false
\def\numberedchapters{\cn@true}
\def\unnumberedchapters{\cn@false\sequentialequations}
\def\chapterreset{\gl@bal\advance\chapternumber by 1
   \ifnum\equanumber<0 \else\gl@bal\equanumber=0\fi
   \sectionnumber=0 \let\sectionlabel=\rel@x
   \ifcn@ \gl@bal\cn@@true {\pr@tect
       \xdef\chapterlabel{\the\chapterstyle{\the\chapternumber}}}%
    \else \gl@bal\cn@@false \gdef\chapterlabel{\rel@x}\fi }
\def\@alpha#1{\count255='140 \advance\count255 by #1\char\count255}
 \def\alphabetic{\n@expand\@alpha}
\def\@Alpha#1{\count255='100 \advance\count255 by #1\char\count255}
 \def\Alphabetic{\n@expand\@Alpha}
\def\@Roman#1{\uppercase\expandafter{\romannumeral #1}}
 \def\Roman{\n@expand\@Roman}
\def\@roman#1{\romannumeral #1}    \def\roman{\n@expand\@roman}
\def\@number#1{\number #1}         \def\Number{\n@expand\@number}
\def\BLANK#1{\rel@x}               
\def\titleparagraphs{\interlinepenalty=9999
     \leftskip=0.03\hsize plus 0.22\hsize minus 0.03\hsize
     \rightskip=\leftskip \parfillskip=0pt
     \hyphenpenalty=9000 \exhyphenpenalty=9000
     \tolerance=9999 \pretolerance=9000
     \spaceskip=0.333em \xspaceskip=0.5em }
\def\titlestyle#1{\par\begingroup \titleparagraphs
     \iftwelv@\fourteenpoint\else\twelvepoint\fi
   \noindent #1\par\endgroup }
\def\spacecheck#1{\dimen@=\pagegoal\advance\dimen@ by -\pagetotal
   \ifdim\dimen@<#1 \ifdim\dimen@>0pt \vfil\break \fi\fi}
\def\chapter#1{\par \penalty-300 \vskip\chapterskip
   \spacecheck\chapterminspace
   \chapterreset \titlestyle{\ifcn@@\chapterlabel.~\fi #1}
   \nobreak\vskip\headskip \penalty 30000
   {\pr@tect\wlog{\string\chapter\space \chapterlabel}} }

\def\section#1{\par \ifnum\lastpenalty=30000\else
   \penalty-200\vskip\sectionskip \spacecheck\sectionminspace\fi
   \gl@bal\advance\sectionnumber by 1
 {\pr@tect
   \xdef\sectionlabel{\ifcn@@ \chapterlabel.\fi
       \the\sectionstyle{\the\sectionnumber}%
                     }%
   \wlog{\string\section\space \sectionlabel}
 }%
   \noindent {\caps\enspace\sectionlabel.~~#1}\par
   \nobreak\vskip\headskip \penalty 30000 }
\def\subsection#1{\par
   \ifnum\the\lastpenalty=30000\else \penalty-100\smallskip \fi
   \noindent\undertext{#1}\enspace \vadjust{\penalty5000}}

\def\undertext#1{\vtop{\hbox{#1}\kern 1pt \hrule}}
\def\ACK{\par\penalty-100\medskip \spacecheck\sectionminspace
   \line{\fourteenrm\hfil ACKNOWLEDGEMENTS\hfil}\nobreak\vskip\headskip }
\def\ack{\subsection{Acknowledgements:}}
\def\APPENDIX#1#2{\par\penalty-300\vskip\chapterskip
   \spacecheck\chapterminspace \chapterreset \xdef\chapterlabel{#1}
   \titlestyle{APPENDIX #2} \nobreak\vskip\headskip \penalty 30000
   \wlog{\string\Appendix~\chapterlabel} }
\def\Appendix#1{\APPENDIX{#1}{#1}}
\def\appendix{\APPENDIX{A}{}}
%
%
%
\def\eqname#1{\rel@x {\pr@tect
  \ifnum\equanumber<0 \xdef#1{{\rm(\number-\equanumber)}}%
     \gl@bal\advance\equanumber by -1
  \else \gl@bal\advance\equanumber by 1
   \xdef#1{{\rm(\ifcn@@ \chapterlabel.\fi \number\equanumber)}}\fi
  }#1}

\def\eqn{\eqno\eqname}

\def\eqinsert#1{\noalign{\dimen@=\prevdepth \nointerlineskip
   \setbox0=\hbox to\displaywidth{\hfil #1}
   \vbox to 0pt{\kern 0.5\baselineskip\hbox{$\!\box0\!$}\vss}
   \prevdepth=\dimen@}}
%

%
%
\def\GENITEM#1;#2{\par \hangafter=0 \hangindent=#1
    \Textindent{$ #2 $}\ignorespaces}
\outer\def\newitem#1=#2;{\gdef#1{\GENITEM #2;}}

\newdimen\itemsize                \itemsize=30pt
\newitem\item=1\itemsize;
\newitem\sitem=1.75\itemsize;     
\newitem\ssitem=2.5\itemsize;     
\outer\def\newlist#1=#2&#3&#4;{\toks0={#2}\toks1={#3}%
   \count255=\escapechar \escapechar=-1
   \alloc@0\list\countdef\insc@unt\listcount     \listcount=0
   \edef#1{\par
      \countdef\listcount=\the\allocationnumber
      \advance\listcount by 1
      \hangafter=0 \hangindent=#4
      \Textindent{\the\toks0{\listcount}\the\toks1}}
   \expandafter\expandafter\expandafter
    \edef\c@t#1{begin}{\par
      \countdef\listcount=\the\allocationnumber \listcount=1
      \hangafter=0 \hangindent=#4
      \Textindent{\the\toks0{\listcount}\the\toks1}}
   \expandafter\expandafter\expandafter
    \edef\c@t#1{con}{\par \hangafter=0 \hangindent=#4 \noindent}
   \escapechar=\count255}
\def\c@t#1#2{\csname\string#1#2\endcsname}
\newlist\point=\Number&.&1.0\itemsize;
\newlist\subpoint=(\alphabetic&)&1.75\itemsize;
\newlist\subsubpoint=(\roman&)&2.5\itemsize;
%

%
%
%
%
\newcount\referencecount     \referencecount=0
\newcount\lastrefsbegincount \lastrefsbegincount=0
\newif\ifreferenceopen       \newwrite\referencewrite
\newdimen\refindent          \refindent=30pt
\def\normalrefmark#1{\attach{\scriptscriptstyle [ #1 ] }}
\let\PRrefmark=\attach
\def\NPrefmark#1{\step@ver{{\;[#1]}}}
\def\refmark#1{\rel@x\ifPhysRev\PRrefmark{#1}\else\normalrefmark{#1}\fi}
\def\refend@{\refmark{\number\referencecount}}
\def\refend{\refend@{}\space }
\def\refsend{\refmark{\count255=\referencecount
   \advance\count255 by-\lastrefsbegincount
   \ifcase\count255 \number\referencecount
   \or \number\lastrefsbegincount,\number\referencecount
   \else \number\lastrefsbegincount-\number\referencecount \fi}\space }
\def\REFNUM#1{\rel@x \gl@bal\advance\referencecount by 1
    \xdef#1{\the\referencecount }}
\def\Refnum#1{\REFNUM #1\refend@ } 
\def\REF#1{\REFNUM #1\R@FWRITE\ignorespaces}
\def\Ref#1{\Refnum #1\REFWRITE }
\def\ref{\Ref\?}
\def\REFS#1{\REFNUM #1\gl@bal\lastrefsbegincount=\referencecount
    \REFWRITE }

\def\r@fitem#1{\par \hangafter=0 \hangindent=\refindent \Textindent{#1}}
\def\refitem#1{\r@fitem{#1.}}
\def\NPrefitem#1{\r@fitem{[#1]}}
\def\NPrefs{\let\refmark=\NPrefmark \let\refitem=\NPrefitem}
\def\REFWRITE{\R@FWRITE\rel@x }
\def\R@FWRITE#1{\ifreferenceopen \else \gl@bal\referenceopentrue
     \immediate\openout\referencewrite=\jobname.refs
     \toks@={\begingroup \refoutspecials \catcode`\^^M=10 }%
     \immediate\write\referencewrite{\the\toks@}\fi
    \immediate\write\referencewrite{\noexpand\refitem %
                                    {\the\referencecount}}%
    \p@rse@ndwrite \referencewrite #1}
\begingroup
 \catcode`\^^M=\active \let^^M=\relax %
 \gdef\p@rse@ndwrite#1#2{\begingroup \catcode`\^^M=12 \newlinechar=`\^^M%
         \chardef\rw@write=#1\sc@nlines#2}%
 \gdef\sc@nlines#1#2{\sc@n@line \g@rbage #2^^M\endsc@n \endgroup #1}%
 \gdef\sc@n@line#1^^M{\expandafter\toks@\expandafter{\deg@rbage #1}%
         \immediate\write\rw@write{\the\toks@}%
         \futurelet\n@xt \sc@ntest }%
\endgroup
\def\sc@ntest{\ifx\n@xt\endsc@n \let\n@xt=\rel@x
       \else \let\n@xt=\sc@n@notherline \fi \n@xt }
\def\sc@n@notherline{\sc@n@line \g@rbage }
\def\deg@rbage#1{}
\let\g@rbage=\relax    \let\endsc@n=\relax
\def\refout{\par\penalty-400\vskip\chapterskip
   \spacecheck\referenceminspace
   \ifreferenceopen \Closeout\referencewrite \referenceopenfalse \fi
   \line{\fourteenrm\hfil REFERENCES\hfil}\vskip\headskip
   \input \jobname.refs
   }
\def\refoutspecials{\sfcode`\.=1000 \interlinepenalty=1000
         \rightskip=\z@ plus 1em minus \z@ }
\def\Closeout#1{\toks0={\par\endgroup}\immediate\write#1{\the\toks0}%
   \immediate\closeout#1}
%
%
\newcount\figurecount     \figurecount=0
\newcount\tablecount      \tablecount=0
\newif\iffigureopen       \newwrite\figurewrite
\newif\iftableopen        \newwrite\tablewrite
\def\FIGNUM#1{\rel@x \gl@bal\advance\figurecount by 1
    \xdef#1{\the\figurecount}}
\def\FIGURE#1{\FIGNUM #1\F@GWRITE\ignorespaces }

\def\figitem#1{\r@fitem{#1)}}
\def\FIGWRITE{\F@GWRITE\rel@x }
\def\TABNUM#1{\rel@x \gl@bal\advance\tablecount by 1
    \xdef#1{\the\tablecount}}
\def\TABLE#1{\TABNUM #1\T@BWRITE\ignorespaces }

\def\tabitem#1{\r@fitem{#1:}}
\def\TABWRITE{\T@BWRITE\rel@x }
\def\F@GWRITE#1{\iffigureopen \else \gl@bal\figureopentrue
     \immediate\openout\figurewrite=\jobname.figs
     \toks@={\begingroup \catcode`\^^M=10 }%
     \immediate\write\figurewrite{\the\toks@}\fi
    \immediate\write\figurewrite{\noexpand\figitem %
                                 {\the\figurecount}}%
    \p@rse@ndwrite \figurewrite #1}
\def\T@BWRITE#1{\iftableopen \else \gl@bal\tableopentrue
     \immediate\openout\tablewrite=\jobname.tabs
     \toks@={\begingroup \catcode`\^^M=10 }%
     \immediate\write\tablewrite{\the\toks@}\fi
    \immediate\write\tablewrite{\noexpand\tabitem %
                                 {\the\tablecount}}%
    \p@rse@ndwrite \tablewrite #1}
\def\figout{\par\penalty-400
   \vskip\chapterskip\spacecheck\referenceminspace
   \iffigureopen \Closeout\figurewrite \figureopenfalse \fi
   \line{\fourteenrm\hfil FIGURE CAPTIONS\hfil}\vskip\headskip
   \input \jobname.figs
   }
\def\tabout{\par\penalty-400
   \vskip\chapterskip\spacecheck\referenceminspace
   \iftableopen \Closeout\tablewrite \tableopenfalse \fi
   \line{\fourteenrm\hfil TABLE CAPTIONS\hfil}\vskip\headskip
   \input \jobname.tabs
   }
%
%
%
\newbox\picturebox
\def\p@cht{\ht\picturebox }
\def\p@cwd{\wd\picturebox }
\def\p@cdp{\dp\picturebox }
\newdimen\xshift
\newdimen\yshift
\newdimen\captionwidth
\newskip\captionskip
\captionskip=15pt plus 5pt minus 3pt
\def\fullwidth{\captionwidth=\hsize }
\newtoks\Caption
\newif\ifcaptioned
\newif\ifselfcaptioned
\def\caption{\captionedtrue \Caption }
\newcount\linesabove
\newif\iffileexists
\newtoks\picfilename
\def\fil@#1 {\fileexiststrue \picfilename={#1}}
\def\file#1{\if=#1\let\n@xt=\fil@ \else \def\n@xt{\fil@ #1}\fi \n@xt }
\def\pl@t{\begingroup \pr@tect
    \setbox\picturebox=\hbox{}\fileexistsfalse
    \let\height=\p@cht \let\width=\p@cwd \let\depth=\p@cdp
    \xshift=\z@ \yshift=\z@ \captionwidth=\z@
    \Caption={}\captionedfalse
    \linesabove =0 \picturedefault }
\def\plot{\pl@t \selfcaptionedfalse }
\def\Picture#1{\gl@bal\advance\figurecount by 1
    \xdef#1{\the\figurecount}\pl@t \selfcaptionedtrue }

\def\s@vepicture{\iffileexists \parsefilename \redopicturebox \fi
   \ifdim\captionwidth>\z@ \else \captionwidth=\p@cwd \fi
   \xdef\lastpicture{\iffileexists
        \setbox0=\hbox{\raise\the\yshift \vbox{%
              \moveright\the\xshift\hbox{\picturedefinition}}}%
        \else \setbox0=\hbox{}\fi
         \ht0=\the\p@cht \wd0=\the\p@cwd \dp0=\the\p@cdp
         \vbox{\hsize=\the\captionwidth \line{\hss\box0 \hss }%
              \ifcaptioned \vskip\the\captionskip \noexpand\Tenpoint
                \ifselfcaptioned Figure~\the\figurecount.\enspace \fi
                \the\Caption \fi }}%
    \endgroup }
\let\endpicture=\s@vepicture
\def\savepicture#1{\s@vepicture \global\let#1=\lastpicture }
\def\displaypicture{\fullwidth \s@vepicture $$\lastpicture $${}}
\def\toppicture{\fullwidth \s@vepicture \topinsert
    \lastpicture \medskip \endinsert }
\def\midpicture{\fullwidth \s@vepicture \midinsert
    \lastpicture \endinsert }
%
%
\def\leftpicture{\pres@tpicture
    \dimen@i=\hsize \advance\dimen@i by -\dimen@ii
    \setbox\picturebox=\hbox to \hsize {\box0 \hss }%
    \wr@paround }
\def\rightpicture{\pres@tpicture
    \dimen@i=\z@
    \setbox\picturebox=\hbox to \hsize {\hss \box0 }%
    \wr@paround }
\def\pres@tpicture{\gl@bal\linesabove=\linesabove
    \s@vepicture \setbox\picturebox=\vbox{
         \kern \linesabove\baselineskip \kern 0.3\baselineskip
         \lastpicture \kern 0.3\baselineskip }%
    \dimen@=\p@cht \dimen@i=\dimen@
    \advance\dimen@i by \pagetotal
    \par \ifdim\dimen@i>\pagegoal \vfil\break \fi
    \dimen@ii=\hsize
    \advance\dimen@ii by -\parindent \advance\dimen@ii by -\p@cwd
    \setbox0=\vbox to\z@{\kern-\baselineskip \unvbox\picturebox \vss }}
\def\wr@paround{\Caption={}\count255=1
    \loop \ifnum \linesabove >0
         \advance\linesabove by -1 \advance\count255 by 1
         \advance\dimen@ by -\baselineskip
         \expandafter\Caption \expandafter{\the\Caption \z@ \hsize }%
      \repeat
    \loop \ifdim \dimen@ >\z@
         \advance\count255 by 1 \advance\dimen@ by -\baselineskip
         \expandafter\Caption \expandafter{%
             \the\Caption \dimen@i \dimen@ii }%
      \repeat
    \edef\n@xt{\parshape=\the\count255 \the\Caption \z@ \hsize }%
    \par\noindent \n@xt \strut \vadjust{\box\picturebox }}
\let\picturedefault=\relax
\let\parsefilename=\relax
\def\redopicturebox{\let\picturedefinition=\rel@x
   \errhelp=\disabledpictures
   \errmessage{This version of TeX cannot handle pictures.  Sorry.}}
\newhelp\disabledpictures
     {You will get a blank box in place of your picture.}
%
%
%
%
%
%
%
%
%
%
\def\FRONTPAGE{\ifvoid255\else\vfill\penalty-20000\fi
   \gl@bal\pagenumber=1     \gl@bal\chapternumber=0
   \gl@bal\equanumber=0     \gl@bal\sectionnumber=0
   \gl@bal\referencecount=0 \gl@bal\figurecount=0
   \gl@bal\tablecount=0     \gl@bal\frontpagetrue
   \gl@bal\lastf@@t=0       \gl@bal\footsymbolcount=0
   \gl@bal\cn@@false }

\def\papers{\papersize\headline=\paperheadline\footline=\paperfootline}
\def\papersize{\hsize=35pc \vsize=50pc \hoffset=0pc \voffset=1pc
   \advance\hoffset by\HOFFSET \advance\voffset by\VOFFSET
   \pagebottomfiller=0pc
   \skip\footins=\bigskipamount \normalspace }
\papers  
%
%
\newskip\lettertopskip       \lettertopskip=20pt plus 50pt
\newskip\letterbottomskip    \letterbottomskip=\z@ plus 100pt
\newskip\signatureskip       \signatureskip=40pt plus 3pt
\def\lettersize{\hsize=6.5in \vsize=8.5in \hoffset=0in \voffset=0.5in
   \advance\hoffset by\HOFFSET \advance\voffset by\VOFFSET
   \pagebottomfiller=\letterbottomskip
   \skip\footins=\smallskipamount \multiply\skip\footins by 3
   \singlespace }
\def\MEMO{\lettersize \headline=\letterheadline \footline={\hfil }%
   \let\rule=\memorule \FRONTPAGE \memohead }

\def\memodate{\afterassignment\MEMO \date }
\def\memit@m#1{\smallskip \hangafter=0 \hangindent=1in
    \Textindent{\caps #1}}
\def\subject{\memit@m{Subject:}}
\def\topic{\memit@m{Topic:}}
\def\from{\memit@m{From:}}
\def\to{\rel@x \ifmmode \rightarrow \else \memit@m{To:}\fi }
\def\memorule{\medskip\hrule height 1pt\bigskip}  
\def\memohead{\centerline{\fourteenrm MEMORANDUM}}
\newwrite\labelswrite
\newtoks\rw@toks
\def\letters{\lettersize
   \headline=\letterheadline \footline=\letterfootline
   \immediate\openout\labelswrite=\jobname.lab}

\let\letterhead=\rel@x
\def\addressee#1{\medskip\line{\hskip 0.75\hsize plus\z@ minus 0.25\hsize
                               \the\date \hfil }%
   \vskip \lettertopskip
   \ialign to\hsize{\strut ##\hfil\tabskip 0pt plus \hsize \crcr #1\crcr}
   \writelabel{#1}\medskip \noindent\hskip -\spaceskip \ignorespaces }
\def\rwl@begin#1\cr{\rw@toks={#1\crcr}\rel@x
   \immediate\write\labelswrite{\the\rw@toks}\futurelet\n@xt\rwl@next}
\def\rwl@next{\ifx\n@xt\rwl@end \let\n@xt=\rel@x
      \else \let\n@xt=\rwl@begin \fi \n@xt}
\let\rwl@end=\rel@x
\def\writelabel#1{\immediate\write\labelswrite{\noexpand\labelbegin}
     \rwl@begin #1\cr\rwl@end
     \immediate\write\labelswrite{\noexpand\labelend}}
\newtoks\FromAddress         \FromAddress={}
\newtoks\sendername          \sendername={}
\newbox\FromLabelBox
\newdimen\labelwidth          \labelwidth=6in
\def\makelabels{\afterassignment\Makelabels \sendername=}
\def\Makelabels{\FRONTPAGE \letterinfo={\hfil } \MakeFromBox
     \immediate\closeout\labelswrite  \input \jobname.lab\vfil\eject}
\let\labelend=\rel@x
\def\labelbegin#1\labelend{\setbox0=\vbox{\ialign{##\hfil\cr #1\crcr}}
     \MakeALabel }
\def\MakeFromBox{\gl@bal\setbox\FromLabelBox=\vbox{\Tenpoint
     \ialign{##\hfil\cr \the\sendername \the\FromAddress \crcr }}}
\def\MakeALabel{\vskip 1pt \hbox{\vrule \vbox{
        \hsize=\labelwidth \hrule\bigskip
        \leftline{\hskip 1\parindent \copy\FromLabelBox}\bigskip
        \centerline{\hfil \box0 } \bigskip \hrule
        }\vrule } \vskip 1pt plus 1fil }
\def\signed#1{\par \nobreak \bigskip \dt@pfalse \begingroup
  \everycr={\noalign{\nobreak
            \ifdt@p\vskip\signatureskip\gl@bal\dt@pfalse\fi }}%
  \tabskip=0.5\hsize plus \z@ minus 0.5\hsize
  \halign to\hsize {\strut ##\hfil\tabskip=\z@ plus 1fil minus \z@\crcr
          \noalign{\gl@bal\dt@ptrue}#1\crcr }%
  \endgroup \bigskip }
\newbox\letterb@x
\def\lettertext{\par \vskip\parskip \unvcopy\letterb@x \par }
\def\multiletter{\setbox\letterb@x=\vbox\bgroup
      \everypar{\vrule height 1\baselineskip depth 0pt width 0pt }
      \singlespace \topskip=\baselineskip }
\def\letterend{\par\egroup}
%
%
%
\newskip\frontpageskip
\newtoks\Pubnum   
\newtoks\Pubtype  \let\pubtype=\Pubtype
\newif\ifp@bblock  \p@bblocktrue
\def\PH@SR@V{\doubl@true \baselineskip=24.1pt plus 0.2pt minus 0.1pt
             \parskip= 3pt plus 2pt minus 1pt }
\def\PHYSREV{\papers\PhysRevtrue\PH@SR@V}

\def\titlepage{\FRONTPAGE\papers\ifPhysRev\PH@SR@V\fi
   \ifp@bblock\p@bblock \else\hrule height\z@ \rel@x \fi }
\def\nopubblock{\p@bblockfalse}

\frontpageskip=12pt plus .5fil minus 2pt
\Pubtype={}
\Pubnum={}
\def\p@bblock{\begingroup \tabskip=\hsize minus \hsize
   \baselineskip=1.5\ht\strutbox \topspace-2\baselineskip
   \halign to\hsize{\strut ##\hfil\tabskip=0pt\crcr
       \the\Pubnum\crcr\the\date\crcr\the\pubtype\crcr}\endgroup}
\def\title#1{\vskip\frontpageskip \titlestyle{#1} \vskip\headskip }
\def\author#1{\vskip\frontpageskip\titlestyle{\twelvecp #1}\nobreak}

\def\address#1{\par\kern 5pt\titlestyle{\twelvepoint\it #1}}
\def\andaddress{\par\kern 5pt \centerline{\sl and} \address}

\def\abstract{\par\dimen@=\prevdepth \hrule height\z@ \prevdepth=\dimen@
   \vskip\frontpageskip\centerline{\fourteenrm ABSTRACT}\vskip\headskip }

%
%
%

\def\\{\rel@x \ifmmode \backslash \else {\tt\char`\\}\fi }
\def\sequentialequations{\rel@x \if\equanumber<0 \else
  \gl@bal\equanumber=-\equanumber \gl@bal\advance\equanumber by -1 \fi }
\def\journal#1&#2(#3){\begingroup \let\journal=\dummyj@urnal
    \unskip, \sl #1\unskip~\bf\ignorespaces #2\rm
    (\afterassignment\j@ur \count255=#3), \endgroup\ignorespaces }
\def\j@ur{\ifnum\count255<100 \advance\count255 by 1900 \fi
          \number\count255 }
\def\dummyj@urnal{%
    \toks@={Reference foul up: nested \journal macros}%
    \errhelp={Your forgot & or ( ) after the last \journal}%
    \errmessage{\the\toks@ }}

\def\topspace{\hrule height 0pt depth 0pt \vskip}

\def\half{\coeff12 }

\def\Buildrel#1\under#2{\mathrel{\mathop{#2}\limits_{#1}}}
\def\becomes#1{\mathchoice{\becomes@\scriptstyle{#1}}
   {\becomes@\scriptstyle{#1}} {\becomes@\scriptscriptstyle{#1}}
   {\becomes@\scriptscriptstyle{#1}}}
\def\becomes@#1#2{\mathrel{\setbox0=\hbox{$\m@th #1{\,#2\,}$}%
        \mathop{\hbox to \wd0 {\rightarrowfill}}\limits_{#2}}}

\def\ket#1{\left| #1\right\rangle}

\def\VEV#1{\left\langle #1\right\rangle}

\let\int=\intop         
\def\lsim{\mathrel{\mathpalette\@versim<}}
\def\gsim{\mathrel{\mathpalette\@versim>}}
\def\@versim#1#2{\vcenter{\offinterlineskip
        \ialign{$\m@th#1\hfil##\hfil$\crcr#2\crcr\sim\crcr } }}
\def\big#1{{\hbox{$\left#1\vbox to 0.85\b@gheight{}\right.\n@space$}}}
\def\Big#1{{\hbox{$\left#1\vbox to 1.15\b@gheight{}\right.\n@space$}}}
\def\bigg#1{{\hbox{$\left#1\vbox to 1.45\b@gheight{}\right.\n@space$}}}
\def\Bigg#1{{\hbox{$\left#1\vbox to 1.75\b@gheight{}\right.\n@space$}}}
%
%
%
%
\let\sec@nt=\sec
\def\sec{\rel@x\ifmmode\let\n@xt=\sec@nt\else\let\n@xt\section\fi\n@xt}
\def\obsolete#1{\message{Macro \string #1 is obsolete.}}
\def\firstsec#1{\obsolete\firstsec \section{#1}}
\def\firstsubsec#1{\obsolete\firstsubsec \subsection{#1}}
\def\thispage#1{\obsolete\thispage \gl@bal\pagenumber=#1\frontpagefalse}
\def\thischapter#1{\obsolete\thischapter \gl@bal\chapternumber=#1}
\def\splitout{\obsolete\splitout\rel@x}
\def\prop{\obsolete\prop \propto }
\def\nextequation#1{\obsolete\nextequation \gl@bal\equanumber=#1
   \ifnum\the\equanumber>0 \gl@bal\advance\equanumber by 1 \fi}
\def\BOXITEM{\afterassigment\B@XITEM\setbox0=}
\def\B@XITEM{\par\hangindent\wd0 \noindent\box0 }
%
%
%
\def\phyzzx{PHY\setbox0=\hbox{Z}\copy0 \kern-0.5\wd0 \box0 X}
        
\everyjob{\xdef\today{\monthname~\number\day, \number\year}
        \input myphyx.tex }
\message{ by V.K.}
\def\underwig#1{{
\setbox0=\hbox{$#1$}
\setbox1=\hbox{}
\wd1=\wd0
\ht1=\ht0
\dp1=\dp0
\setbox2=\hbox{$\rm\widetilde{\box1}$}
\dimen@=\ht2 \advance \dimen@ by \dp2 \advance \dimen@ by 1.5pt
\ht2=0pt \dp2=0pt
\hbox to 0pt{$#1$\hss} \lower\dimen@\box2
}}
\def\bunderwig#1{{
\setbox0=\hbox{$#1$}
\setbox1=\hbox{}
\wd1=\wd0
\ht1=\ht0
\dp1=\dp0
\setbox2=\hbox{$\seventeenrm\widetilde{\box1}$}
\dimen@=\the\ht2 \advance \dimen@ by \the\dp2 \advance \dimen@ by 1.5pt
\ht2=0pt \dp2=0pt
\hbox to 0pt{$#1$\hss} \lower\dimen@\box2
}}
\def\ack{\ACK}   
\font\titlerm=cmr10 scaled \magstep 4
\def\TITLEPAGE{\frontpagetrue\pageno=1\pagenumber=1}
\def\WISCONSIN{\vskip15pt\vbox{\hbox{\centerline{\it Department of Physics}}
        \vskip 0pt
  \hbox{\centerline{\it University of Wisconsin, Madison, WI 53706 USA}}}}
\def\TITLE#1{\vskip 1in \centerline{\titlerm #1}}
\def\MORETITLE#1{\vskip 19pt \centerline{\titlerm #1}}
\def\AUTHOR#1{\vskip .5in \centerline{#1}}

\def\ABSTRACT#1{\vskip .5in \vfil \centerline{\twelvepoint \bf Abstract}
        #1 \vfil}
\def\ENDTITLEPAGE{\vfill\eject\pageno=2\pagenumber=2}
\let\letterhead=\MADPHHEAD

\def\semi{;\hfil\break}
\def\sqr#1#2{{\vcenter{\hrule height.#2pt
      \hbox{\vrule width.#2pt height#1pt \kern#1pt
        \vrule width.#2pt}
      \hrule height.#2pt}}}
\def\rect#1#2#3#4{{\vcenter{\hrule height#3pt
      \hbox{\vrule width#4pt height#1pt \kern#1pt
        \vrule width#4pt}
      \hrule height#3pt}}}

\def\bx{{\vcenter{\hrule height 0.4pt
      \hbox{\vrule width 0.4pt height 10pt \kern 10pt
        \vrule width 0.4pt}
      \hrule height 0.4pt}}}
\def\inner{\,{\vcenter{
      \hbox{ \kern 4pt
        \vrule width 0.5pt height 7pt}
      \hrule height 0.5pt}}\,}
\def\square{\mathchoice\sqr34\sqr34\sqr{2.1}3\sqr{1.5}3}
\def\to{{\,\rightarrow\,}}
\def\up#1{\leavevmode \raise.16ex\hbox{#1}}
%
%
%
%
%

%
\newdimen\madphheadsize \madphheadsize=7.0in
\newdimen\madphheadleft \madphheadleft=3.75in
\newdimen\madpheadsize \madpheadsize=7.0in
\newdimen\madpheadleft \madpheadleft=3.5in
\def\MEMO{\letterstyle\FRONTPAGE \letterfrontheadline={\hfil}
    \line{\quad\fourteenss MEMORANDUM\hfil\twelvess\the\date\quad}
    \medskip \memod@f}
\def\MEMOT{\letterstyle\FRONTPAGE \MADAFHEAD\letterfrontheadline={\hfil}
    \line{\quad\fourteenss MEMORANDUM\hfil\twelvess\the\date\quad}
    \medskip \memod@f}
\def\MEMOX{\letterstyle\FRONTPAGE \MADPHHEAD\letterfrontheadline={\hfil}
    \line{\quad\fourteenss MEMORANDUM\hfil\twelvess\the\date\quad}
    \medskip \memod@f}
\showboxbreadth=1000 %
\showboxdepth=5
\newdimen\madphheadsize \madphheadsize=7.0in
\newdimen\madphheadleft \madphheadleft=3.6875in
\def\journal#1&#2,#3(#4){\unskip, \sl #1~\bf #2,\rm  #3 (19#4)}
\def\dollarsigns#1{\ifmmode #1 \else $#1 $ \fi}

%
%
%
\def\figitem#1{\r@fitem{#1.}}
\def\tabitem#1{\r@fitem{#1.}}

\def\sequentialequations{\rel@x \ifnum\equanumber<0 \else
  \gl@bal\equanumber=-\equanumber \gl@bal\advance\equanumber by -1 \fi }

\def\boxit#1{\vbox{\hrule\hbox{\vrule\kern3pt
\vbox{\kern3pt#1\kern3pt}\kern3pt\vrule}\hrule}}
%
%
%
%
%
\newif\ifpr@printstyle \pr@printstylefalse
\newbox\leftpage \newdimen\fullhsize \newdimen\hstitle \newdimen\hsbody
\def\preprintstyle{%
       \message{(This will be printed PREPRINTSTYLE)} \let\lr=L
       \frontpagetrue
       \pr@printstyletrue
       \vsize=7truein
       \hsbody=4.75truein
       \fullhsize=10truein
       \hstitle=8truein
       \normalspace
       \Tenpoint
       \voffset=-.31truein
       \hoffset=-.46truein
       \iffrontpage\hsize=\hstitle\else\hsize=\hsbody\fi
 \output={%
    \iffrontpage
      \shipout\vbox{\special{\printertype}\makeheadline
      \hbox to \fullhsize{\hfill\pagebody\hfill}}
      \advancepageno
    \else
       \almostshipout{\leftline{\vbox{\pagebody\makefootline}}}\advancepageno
    \fi}
        \def\almostshipout##1{\if L\lr \count2=1
             \message{[\the\count0.\the\count1.\the\count2]}
        \global\setbox\leftpage=##1 \global\let\lr=R
                             \else \count2=2
        \shipout\vbox{\special{\printertype}
        \hbox to\fullhsize{\hfill\box\leftpage\hskip0.5truein##1\hfill}}
        \global\let\lr=L     \fi}
   \multiply\chapterminspace by 7 \divide\chapterminspace by 9
   \multiply\sectionminspace by 7 \divide\sectionminspace by 9
   \multiply\referenceminspace by 7 \divide\referenceminspace  by 9
   \multiply\chapterskip by 7 \divide\chapterskip  by 9
   \multiply\sectionskip  by 7 \divide\sectionskip  by 9
   \multiply\headskip   by 7 \divide\headskip by 9
   \multiply\baselineskip   by 7 \divide\baselineskip by 9
   \multiply\abovedisplayskip by 7 \divide\abovedisplayskip by 9
   \belowdisplayskip = \abovedisplayskip
}

\tolerance=1000
\def\printertype{ps: }
       \hsbody=6truein
       \fullhsize=6truein
       \hstitle=6truein
       \normalspace
       \Twelvepoint
       \hoffset=0.3truein
       \voffset=0.2truein
       \hsize=\hsbody
\paperheadline={\ifdr@ftmode\hfil\draftdate\else\hfill\fi}
\def\advancepageno{\gl@bal\advance\pagen@ by 1
   \ifnum\pagenumber<0 \gl@bal\advance\pagenumber by -1
    \else\gl@bal\advance\pagenumber by 1 \fi
    \gl@bal\frontpagefalse  \swing@
    \gl@bal\hsize=\hsbody} 
\def\papersize{}
\papers
\def\lettersize{ \fullhsize=6.5in \hsbody=6.5in
      \hsize=\fullhsize \vsize=8.5in \hoffset=0in \voffset=0.5in
      \advance\hoffset by\HOFFSET \advance\voffset by\VOFFSET
      \pagebottomfiller=\letterbottomskip
   \skip\footins=\smallskipamount \multiply\skip\footins by 3
   \singlespace }
\def\semi{;\hfil\break}
%
%
\newtoks\chapterheadstyle  \chapterheadstyle={\relax}
\def\chapter#1{{\the\chapterheadstyle\par \penalty-300 \vskip\chapterskip
   \spacecheck\chapterminspace
   \chapterreset \titlestyle{\ifcn@@\chapterlabel.~\fi #1}
   \nobreak\vskip\headskip \penalty 30000
   {\pr@tect\wlog{\string\chapter\space \chapterlabel}} }}

\def\APPENDIX#1#2{{\the\chapterheadstyle\par\penalty-300\vskip\chapterskip
   \spacecheck\chapterminspace \chapterreset \xdef\chapterlabel{#1}
   \titlestyle{APPENDIX #2} \nobreak\vskip\headskip \penalty 30000
   \wlog{\string\Appendix~\chapterlabel} }}
\def\chapterreset{\gl@bal\advance\chapternumber by 1
   \ifnum\equanumber<0 \else\gl@bal\equanumber=0\fi
   \gl@bal\sectionnumber=0 \let\sectionlabel=\rel@x
   \ifcn@ \gl@bal\cn@@true {\pr@tect
       \xdef\chapterlabel{{\the\chapterstyle{\the\chapternumber}}}}%
    \else \gl@bal\cn@@false \gdef\chapterlabel{\rel@x}\fi }
%
%
\newif\ifdr@ftmode
\def\draftmode{
   \tenpoint
   \baselineskip=24pt plus 2pt minus 2pt
   \dr@ftmodetrue
   \message{ DRAFTMODE }
   \writelabels
   \def\timestring{\begingroup
     \count0 = \time \divide\count0 by 60
     \count2 = \count0  
     \count4 = \time \multiply\count0 by 60
     \advance\count4 by -\count0   
     \ifnum\count4<10 \toks1={0} 
     \else \toks1 = {}
     \fi
     \ifnum\count2<12 \toks0={a.m.} %
          \ifnum\count2<1 \count2=12 \fi
     \else            \toks0={p.m.} %
           \ifnum\count2=12 
           \else
           \advance\count2 by -12 
           \fi
     \fi
     \number\count2:\the\toks1 \number\count4\thinspace \the\toks0
   \endgroup}%
   \def\draftdate{{{\tt preliminary version:}\space{\rm
                                  \timestring\quad\the\date}}}
}
\def\nolabels{\def\leqlabel##1{}\def\eqlabel##1{}\def\reflabel##1{}%
\def\leqlabel##1{}}
\def\writelabels{
\def\eqlabel##1{{\escapechar-1\rlap{\sevenrm\hskip.05in\string##1}}}%
\def\leqlabel##1{{\escapechar-1\llap{\sevenrm\string##1\hskip.05in}}}%
\def\reflabel##1{\escapechar-1\ ``{\tt\string##1}"\ }  }

\nolabels        
\dr@ftmodefalse  
%
%
%

\def\eqn#1{\eqno\eqname{#1}\eqlabel#1}
%
%
\def\eqinsert#1{\noalign{\dimen@=\prevdepth \nointerlineskip
   \setbox0=\hbox to\displaywidth{\hfil #1}
   \vbox to 0pt{\kern 0.5\baselineskip\hbox{$\!\box0\!$}\vss}
   \prevdepth=\dimen@}}  

%


%
%
%
%
%
\def\REF#1#2{\REFNUM #1\REFWRITE{\ignorespaces \reflabel#1 #2}}
\def\Ref#1#2{\Refnum #1\REFWRITE{\reflabel#1 #2}}
\def\REFS#1#2{\REFNUM #1\gl@bal\lastrefsbegincount=\referencecount
\REFWRITE{\reflabel#1 #2}}

\def\refout{\par\penalty-400\vskip\chapterskip
   \spacecheck\referenceminspace
   \ifreferenceopen \Closeout\referencewrite \referenceopenfalse \fi
   \line{\ifpr@printstyle\twelverm\else\fourteenrm\fi
         \hfil REFERENCES\hfil}\vskip\headskip
   \input \jobname.refs
   }
\def\ACK{\par\penalty-100\medskip \spacecheck\sectionminspace
   \line{\ifpr@printstyle\twelverm\else\fourteenrm\fi
      \hfil ACKNOWLEDGEMENTS\hfil}\nobreak\vskip\headskip }
\def\tabout{\par\penalty-400
   \vskip\chapterskip\spacecheck\referenceminspace
   \iftableopen \Closeout\tablewrite \tableopenfalse \fi
   \line{\ifpr@printstyle\twelverm\else\fourteenrm\fi\hfil TABLE CAPTIONS\hfil}
   \vskip\headskip
   \input \jobname.tabs
   }
\def\figout{\par\penalty-400
   \vskip\chapterskip\spacecheck\referenceminspace
   \iffigureopen \Closeout\figurewrite \figureopenfalse \fi
   \line{\ifpr@printstyle\twelverm\else\fourteenrm\fi\hfil FIGURE
CAPTIONS\hfil}
   \vskip\headskip
   \input \jobname.figs
   }
\def\masterreset{\begingroup\hsize=\hsbody
   \global\pagenumber=1 \global\chapternumber=0
   \global\equanumber=0 \global\sectionnumber=0
   \global\referencecount=0 \global\figurecount=0 \global\tablecount=0
   \endgroup}
%

%
\def\half{{\textstyle{1\over2}}}

\def\12{{1\over2}}

\def\sla{\raise.15ex\hbox{$/$}\kern-.57em}
\def\leaderfill{\leaders\hbox to 1em{\hss.\hss}\hfill}
\def\dual{{\,^*\kern-.20em}}
\def\bx{{\vcenter{\hrule height 0.4pt
      \hbox{\vrule width 0.4pt height 10pt \kern 10pt
        \vrule width 0.4pt}
      \hrule height 0.4pt}}}
\def\inner{\,{\vcenter{
      \hbox{ \kern 4pt
        \vrule width 0.5pt height 7pt}
      \hrule height 0.5pt}}\,}
\def\twiddle{\lower.9ex\rlap{$\kern-.1em\scriptstyle\sim$}}
\def\bigtwiddle{\lower1.ex\rlap{$\sim$}}
\def\gtwid{\mathrel{\raise.3ex\hbox{$>$\kern-.75em\lower1ex\hbox{$\sim$}}}}
\def\ltwid{\mathrel{\raise.3ex\hbox{$<$\kern-.75em\lower1ex\hbox{$\sim$}}}}
\def\square{\kern1pt\vbox{\hrule height 1.2pt\hbox{\vrule width 1.2pt\hskip 3pt
   \vbox{\vskip 6pt}\hskip 3pt\vrule width 0.6pt}\hrule height 0.6pt}\kern1pt}
\def\tdot#1{\mathord{\mathop{#1}\limits^{\kern2pt\ldots}}}

\def\pmb#1{\setbox0=\hbox{#1}    
  \kern-.025em\copy0\kern-\wd0
  \kern  .05em\copy0\kern-\wd0
  \kern-.025em\raise.0433em\box0 }

\def\grad{\nabla}
\def\const{{\rm const}}

\def\pr{\journal Phys. Rev. }

\def\prl{\journal Phys. Rev. Lett. }

\hyphenation{anom-aly anom-alies coun-ter-term coun-ter-terms}
\def\inv{^{\raise.15ex\hbox{${\scriptscriptstyle -}$}\kern-.05em 1}}

\def\Dsl{\,\raise.15ex\hbox{/}\mkern-13.5mu D} 
\def\dsl{\raise.15ex\hbox{/}\kern-.57em\partial}

\def\boxeqn#1{\vcenter{\vbox{\hrule\hbox{\vrule\kern3pt\vbox{\kern3pt
        \hbox{${\displaystyle #1}$}\kern3pt}\kern3pt\vrule}\hrule}}}
\def\mbox#1#2{\vcenter{\hrule \hbox{\vrule height#2in
                \kern#1in \vrule} \hrule}}  
%

\def\grad#1{\,\nabla\!_{{#1}}\,}

\def\psibar{\overline\psi}

\def\darr#1{\raise1.5ex\hbox{$\leftrightarrow$}\mkern-16.5mu #1}
\def\roughly#1{\raise.3ex\hbox{$#1$\kern-.75em\lower1ex\hbox{$\sim$}}}
%
%
%
%
\def\journal#1&#2(#3){\unskip, \sl #1~\bf #2 \rm (19#3) }
\def\npjournal#1&#2&#3&#4&{\unskip, #1~\rm #2 \rm (#3) #4}
\gdef\prjournal#1&#2&#3&#4&{\unskip, #1~\bf #2, \rm #4 (#3)}

\def\ket#1{\left| #1\right\rangle}
\def\VEV#1{\left\langle #1\right\rangle}

\let\int=\intop         
\catcode`\@=12 
\masterreset
\normalspace
\chapterheadstyle={\bf}
\overfullrule=0pt
\def\pr{{\it Phys. Rev.\/ }}
\def\prl{{\it Phys. Rev. Lett.\/ }}
\def\psisquare{{|\psi|^2}}
\def\grad{\nabla}
\def\psibar{\overline{\psi}}
\def\parbar{\overline{\partial}}
\def\zbar{{\overline{z}}}
\def\Ath{$A^{\rm th}$\ }

\TITLEPAGE
\rightline{{\tenpoint\baselineskip=12pt
           \vtop{\hbox{\strut MAD/TH-92-02}
                 \hbox{\strut hep-th/9206073}
                 \hbox{\strut April 1992}}}}
\TITLE{ Charged Vortex Dynamics in Ginzburg-Landau Theory }
\MORETITLE{ of the Fractional Quantum Hall Effect }
\AUTHOR{ Theodore J. Allen{${}^*$} {\it and} Andrew J. Bordner{${}^{\dag}$} }
\footnote{*}{tjallen@wishep.physics.wisc.edu}
\footnote{\dagger}{bordner@wishep.physics.wisc.edu}
\WISCONSIN

\ABSTRACT{ We write a Ginzburg-Landau Hamiltonian for a charged order parameter
interacting with a background electromagnetic field in 2+1 dimensions.  Using
the method of Lund we derive a collective coordinate action for vortex defects
in the order parameter and demonstrate that the vortices are charged. We
examine the classical dynamics of the vortices and then quantize their motion,
demonstrating that their peculiar classical motion is a result of the fact that
the quantum motion takes place in the lowest Landau level.  The classical and
quantum motion in two dimensional regions with boundaries is also investigated.
The quantum theory is not invariant under magnetic translations. Magnetic
translations add total time derivative terms to the collective action, but no
extra constants of the motion result. }

\ENDTITLEPAGE

\chapter{ Introduction }

The microscopic physics of the fractional quantum Hall state is by now
reasonably well understood.\Ref\FQHE{ {\sl The Quantum Hall Effect},
R.E.~Prange and S.M.~Girvin, eds., (Springer, Berlin, Heidelberg, New York,
1987), and references therein.} When the density of electrons is one per an
area through which an  odd number, $2n+1$, of flux quanta pass, there can be
quasiparticle excitations in the many-body state which cost little  energy to
produce and which have a fractional effective charge
$e/(2n+1)$.\Ref\Laughlin{R.B.~Laughlin, \prl {\bf 50} (1983) 1395\semi
B.I.~Halperin, \prl {\bf 52} (1984) 1583\semi D.~Arovas, R.~Schrieffer and
F.~Wilczek, \prl {\bf 53} (1984) 722\semi R.~Tao and Y.S.~Wu, \pr {\bf B31}
(1985) 6859\semi D.J.~Thouless and Y.S.~Wu, \pr {\bf B31} (1985) 1191\semi
D.A.~Arovas, R.~Schrieffer and A.~Zee, {\it Nucl. Phys.} {\bf B251} (1985)
117.} These excitations are described as vortices in the Laughlin wave
function.  The fractionally charged quasiparticles themselves form an extended
state and can transport charge across a sample.  Finally, a many-quasiparticle
state may itself have vortex excitations, and so on, from which the whole
hierarchy\Ref\Haldane{F.D.M.~Haldane, \prl {\bf 51} (1983) 605\semi
B.I.~Halperin, \prl {\bf 52} (1984) 1583; {\it Ibid.\/} 2390 (E)\semi
R.B.~Laughlin, {\it Surface Science} {\bf 141} (1984) 11.} of fractional
conductivity states may be formed.

While the microscopic picture is compelling, there is not yet a complete
macroscopic phenomenological description, such as a Ginzburg-Landau
theory,\Ref\GinzLan{V.I. Ginzburg and L.D. Landau, {\it Zh. Eksp. Teor. Fiz.\/}
{\bf 20} (1950) 1937.} for the fractional quantum Hall system.  At present one
must put in by hand such features as the fractional filling. Ginzburg-Landau
theories are useful for describing many-body systems such as liquid Helium and
superconductivity, and one might hope that they would find application to the
fractional quantum Hall effect as well.  It is well-known that Ginzburg-Landau
theories can  describe vortex excitations, both in liquid Helium, where they
are true macroscopic fluid vortices,\Ref\Feynman{L. Onsager, {\it Nuovo
Cimento} {\bf 6} Suppl. 2 (1949) 249\semi R.P. Feynman, {\sl Prog. of Low Temp.
Physics}, C.J.~Gorter, Ed., Vol.\ I (North-Holland, Amsterdam, 1955) p.\ 17.}
and in type II superconductors, where they are macroscopic lines of magnetic
flux.\Ref\Abrikosov{A.A.~Abrikosov, {\it Zh. Eksp. Teor. Fiz.\/} {\bf 32}
(1957) 1442; {\it Sov. Phys. JETP\/} {\bf 5} (1957) 1174.} In both of these
cases, the vortex excitations, which are uncharged, may  be observed directly.
Vortex solutions, known as cosmic strings or Nielsen-Olesen\Ref\NO{H. Nielsen
and P. Olesen, {\it Nucl. Phys.\/} {\bf B61} (1973) 45.} strings,  are also
postulated to exist in fundamental field theories of particle physics.  There
the theory is derived not from some phenomenological Landau free energy, but
from the fundamental Lagrangian of the Higgs fields necessary for symmetry
breaking and fermion mass generation.\REF\Davis{R.L. Davis, {\it Mod. Phys.
Lett.\/} {\bf A5} (1990) 853.} Although Ginzburg-Landau vortices and
Nielsen-Olesen vortices are quite similar, there are significant differences in
their dynamics.\refmark{\Davis}  We wish to extend the vortex description to
fractional quantum Hall systems.  We note that this idea has been advocated
many places in the literature.\Ref\GLIdea{S.~Girvin and A.~MacDonald, \prl {\bf
58} (1987) 1252\semi S.M.~Girvin, {\it in ref.\/}  \FQHE, p.\ 389\semi
S.C.~Zhang, T.H.~Hansson and S.~Kivelson, \prl {\bf 62}  (1989) 82\semi
N.~Read, \prl {\bf 62} (1989) 86.}

In Sec.\ 2 we write the most general scalar gauge-invariant Landau free energy
function which is second order in spatial derivatives.  We fix the parameters
in order to find a solution which is a reasonable approximation of a  uniform
charge density.  In this charged uniform fluid we postulate vortex-like defect
solutions.  In Sec.\ 3, we use the method of Lund\Ref\Lund{F. Lund, \it Phys.
Lett.  \bf A 159 \rm (1991) 245.} to find the collective coordinate  action for
the motion of the vortices and demonstrate that they are charged and
act as though they carry a single unit of magnetic flux.

The equations of motion for the system appear to be classical, but are best
interpreted quantum mechanically.  In Sec.\ 4 we apply the method of Dirac to
the quantization of the collective coordinate action. The vortices may be
quantized as having any statistics we like, by postulating the behavior of the
many-body wave function under an exchange of two vortices.  These arbitrary
statistics might be introduced through the use of non-dynamical Chern-Simons
fields if the order parameter field is to be single-valued.  We also consider
in Sec.\ 4 the motion of vortices in samples with simple boundaries, and find
that questions of gauge are resolved by the boundary conditions. In  Sec.\ 5 we
show that magnetic translations are not a symmetry of the quantum  theory
because they add total time derivative terms, sometimes called theta terms, to
the effective action.

\chapter{ The Effective Hamiltonian }

In the theory of superconductivity, a Landau free energy is postulated for an
order parameter which has the kinetic term of a charged boson. The vector
potential is not the external vector potential alone, but includes a
dynamically determined piece.   In order to minimize the free energy,  the
vector potential within most of the superconducting material must be curl-free,
thus the magnetic field is concentrated in vortex lines where the material is
not superconducting.  To observe the fractional quantum Hall effect, the
magnetic fields must be extremely large.   Because of the strength of the
magnetic field, we do not expect the microscopic field to be bundled into
vortices and vanish throughout most of the sample. It seems quite reasonable,
therefore, to neglect the self-generated vector potential in the Landau free
energy.

\REF\Karabali{D. Karabali and B. Sakita, {\it Int. J. Mod. Phys.\/} {\bf A6}
(1991) 5079\semi D. Karabali and B. Sakita, to appear in the proceedings of the
International Sakharov conference, Moscow 1991\semi D. Karabali, {\it
Collective Field Representation of Nonrelativistic Fermions in (1+1)
Dimensions\/}, City College preprint CCNY-HEP-91/11, hepth@xxx/9109013.} We
begin by writing the most general gauge invariant scalar free energy (or
Hamiltonian) function  of a complex order parameter $\psi$, subject to an
external magnetic field.  In a superconductor, the order parameter is
essentially the wave function of a Cooper pair.  The interpretation of the
order parameter for the fractional quantum Hall system from microscopic
variables is not known, but might be, for instance, a collective field for the
charge density.  An example of collective field theory, for fermions in
$1+1$ dimensions, is constructed in ref.\ [\Karabali].

The most general free energy function involving at most two derivatives is
$$\eqalign{{\cal H}_{GL}(\psi,\psibar) &= f_1(\psisquare)\overline{D_a\psi}
D_a\psi + i\,f_2(\psisquare)\epsilon_{ab}\overline{D_a\psi}D_b\psi \cr
&\phantom{=}\,+ {\rm Re}\left[f_3(\psisquare)\psibar D_a\psi\psibar
D_a\psi\right] + {\rm Im}\left[f_4(\psisquare)\psibar D_a\psi\psibar
D_a\psi\right] \cr &\phantom{=}\,+ V(\psisquare) +
i\,\partial_a\left({k\over\psisquare}
\epsilon_{ab}\overline{\psi}\partial_b\psi\right).\cr} \eqn\genH$$
Here $D_a = \partial_a - ieA^{\rm ext}_a$ is the gauge covariant derivative,
and $k$ is a constant.

Terms of higher order in derivatives are irrelevant for critical phenomena,
and can be neglected.  The last term above is actually a surface term which
measures the total change in the phase of $\psi$ around the boundary of the
sample.  The total change in phase is a gauge invariant quantity which will
count the total `vorticity' of the configuration.  We write the most general
free energy \genH\ for completeness and later on we will specialize the free
energy by choosing the simplest functions $f_i(\rho)$ guaranteeing vortex
excitations.

We have avoided introducing Chern-Simons gauge fields {\it ab initio},
preferring instead that they come out as derived quantities.  Even though
Chern-Simons gauge fields have not been explicitly introduced, we find in Sec.\
4 that charged vortices do act as though they carry a flux in addition to their
charge, which is suggestive of Chern-Simons type interactions and fractional
statistics.  However, we find in Sec.\ 4.3 that canonical  quantization
allows any type of statistics.

The non-linear Schr\"odinger equation, $i\dot\psi(x) = \delta {\cal
H}_{GL}/\delta\bar\psi(x)$ with Hamiltonian function \genH, implies the
following equations of motion for the phase, $\phi$, and the square modulus,
$\rho$, of the order parameter $\psi = \sqrt{\rho}e^{i\phi}$.
$$\eqalign{\dot\rho &= -2\left[\rho f_1-\rho^2f_3\right]  \grad\cdot(\nabla
\phi - e{\bf A})\cr &\phantom{=}\,-2\left[f_1 + \rho f_1' - \rho^2f_3'  - 2\rho
f_3\right] \grad\rho\cdot(\grad\phi - e{\bf A})\cr  &\phantom{=}\,- \rho
f_4\grad^2\rho - \left[\rho f_4' + f_4\right] (\grad\rho)^2, \cr &\cr
\dot\phi &= f_1{\grad^2\rho\over 2\rho} + \half(\rho f_1')\left({
\grad\rho\over\rho}\right)^2 - (f_1 + \rho
f_1')\left[\left({\grad\rho\over2\rho}\right)^2 +  \left(\grad\phi - e{\bf A}
\right)^2\right]\cr  &\phantom{=}\,  + (2\rho f_3 + \rho^2 f_3')(\grad\phi -
e{\bf A})^2 + \half f_3\grad^2\rho + (\half f_3')(\grad\rho)^2\cr
&\phantom{=}\,  +  ef_2\epsilon_{ab}\partial_aA_b  - V^{\prime}(\rho).\cr}
\eqn\psimotion$$
We look for solutions with constant $\rho= \rho_0$ and constant $\dot\phi$,
because they will be a good analogy to a fluid in which vortices can form.   It
is easy to see that in general there can be no such solution if $\bf A$ has a
non-zero curl, unless the functions $f_i(\rho)$ are chosen carefully.  For now
we assume that there is some non-negative function $f_1(\rho)$ such that  at
the minima, $\rho = \rho_0$, of $V(\rho)$, the following conditions are
satisfied:
$$\rho_0 f_1'(\rho_0) + f_1(\rho_0) = 0,\quad \rho_0 f_1(\rho_0) = \alpha,
\quad {\rm and}\,\, f_1(0) < \infty ,\eqn\fcond$$
where $\alpha$ is some positive constant.  The last condition is necessary  in
order that vortices, which have $(\grad\phi)^2$ singular at their cores, can
avoid making an infinite contribution to the free energy \genH\ by having
$\rho$ vanish there. The function $f_1(\psisquare) = \alpha/\psisquare$, is an
example meeting all conditions but the last.  For now it is a reasonable
assumption because we are going to ignore the problem of the singularity at the
core when deriving the collective dynamics of the vortices. It is simplest to
take the functions $f_2$, $f_3$ and $f_4$ to be zero.   As an aside we note
that an interesting ansatz for the function $f_1(\rho)$ is
$$ f_1(\rho) = \left[1-\left({\rho\over\pi C}\right)\sin
\left({\pi{C}\over\rho}\right)\right],\eqn\fansatz$$
which has the properties \fcond\ above when $\rho_0 = {C\over 2n+1}$.

In a Coulomb gauge, $\grad\cdot{\bf A} = 0$,  the equations of motion
\psimotion\ reduce to the Laplace equation $\grad^2\phi =0$ for the phase
function $\phi(x)$.  Thus any single vortex solutions, $\phi = \phi_A(x)$,
may be superposed to form a multi-vortex solution $\phi = \sum_A \phi_A(x)$.

\chapter{ Collective Coordinate Action }

In a beautiful paper,\refmark\Lund Lund has given a construction of the
equations of defect dynamics\Ref\defect{A.L.~Fetter, \pr {\bf 151} (1966)
100\semi J.~Creswick and N.~Morrison, {\it Phys. Lett.} {\bf A76} (1980)
276\semi F.D.M. Haldane and Yong-Shi Wu, \prl {\bf 55} (1985) 2887\semi
H.~Kuratsuji, \prl {\bf 68} (1992) 1746.} from a variational principle.  His
method is to substitute the  collective coordinate vortex ansatz into the
action and to vary that action with respect to the collective coordinates.   We
now show that Lund's method also works to find the dynamics of charged
vortices.  We follow closely the argument, and use the notation, in ref.\
[\Lund].  With the assumptions of constant density, $|\psi|^2 = \rho_0$,  all
the $f_i(\rho)$ being zero except $f_1(\rho)$, which must satisfy \fcond,  the
Schr\"odinger action
$$S = \int dt\,d^{2}\!x\,\left[ i\,\bar\psi\dot\psi - {\cal H}_{GL}
(\psi,\bar\psi)\right], \eqn\fieldaction$$
reduces to
$$S_\phi = - \int dt\,d^{2}\!x\,\left[\rho_0\dot\phi+\alpha(\grad\phi-e{\bf
A})^2\right].\eqn\phiaction$$
The phase of the  $N$ vortex order parameter is explicitly given by the
multi-vortex ansatz
$$\phi({\bf x}, t) = \sum_{A=1}^N n_A \phi_A({\bf x}, {\bf X}^A(t)),
\eqn\multivor $$
where the function $\phi_A({\bf x}, {\bf X}^A(t))$ is the polar angle  of the
difference vector $\bf x - {\bf X}^A(t)$ from the center of the  \Ath vortex to
the field point $\bf x$,
$$\phi_A({\bf x}, {\bf X}^A(t)) = \tan^{-1}\left({x_2 - X_2^A(t)\over  x_1 -
X_1^A(t)}\right),\eqn\phiA$$
and the integer $n_A$ is the winding number or `vorticity' of the \Ath vortex.
Because $\phi$ is a multivalued  function,  it is useful to take the domain of
integration to have cuts in it in order  that all quantities in the integrand
be single-valued.  We also take the region of integration to exclude the cores
of the vortices.  Figure 1 shows the integration region and boundary cuts for a
two vortex configuration.  After  substitution of the ansatz \multivor\ into
\phiaction, the first term reduces to a sum of integrals over the cuts  $S_A =
\{ {\bf x}(t) = {\bf Y}^A(s,t)|\, 0< s < \infty\}$ from the \Ath vortex to
infinity,  each of the form:
$$\eqalign{-n_A\int d^{2}\!x\,dt\,\rho_0\partial_t\phi&=-n_A\rho_0\int dt\,{\sl
d\over dt}\int d^2\!x\,\phi_A +2\pi n_A\rho_0\int_{S_A}
d\ell_a\epsilon_{ab}V_b\cr &=2\pi n_A\rho_0\int dt\int_0^\infty
ds\,\epsilon_{ab} {\partial Y_a\over\partial s}{\partial Y_b\over\partial t}\cr
&= \pi n_A\rho_0\int dt\, \epsilon_{ab}X^A_a{\dot X}^A_b.\cr}\eqn\phidot$$
The last equality is obtained by two partial integrations.  For now we ignore
the total derivative term, which is irrelevant for classical mechanics, but we
must return to the question of adding total time derivatives when quantizing
the system. The second term in \phiaction\ leads to a double sum over $A$ and
$B$ of terms of the form
$$\eqalign{-n_An_B\alpha\int d^2\!x\,dt\,\left[\partial_a\phi_A\partial_a\phi_B
\right] &=-n_An_B\alpha\int dt\left(\int_{S_+} + \int_{S_-} + \int_{S_T}  +
\int_{S_\infty}\right)d\ell_a\epsilon_{ab}\phi_A\partial_b\phi_B\cr &= -2\pi
n_An_B\alpha\int dt\int_{S_A}d\ell_a\epsilon_{ab}\partial_b\phi_B\cr &= -2\pi
n_An_B\alpha\int dt\, \ln|{\bf X}^A - {\bf X}^B|,\cr}\eqn\dphidphi$$
and a sum over $B$ of the terms
$$\eqalign{2e\alpha n_B \int dt\,d^2\!x\,(A_a\partial_a\phi_B )&=4\pi e\alpha
n_B \int dt\int_{S_B}d\ell_a\epsilon_{ab}A_b(x)\cr &= 4\pi e\alpha n_B \int
dt\int^\infty_0 ds\,{\partial Y_a\over\partial s} \epsilon_{ab} A_b({\bf Y})\cr
&= 4\pi e\alpha n_B \int dt\int^\infty_0 ds\,{\partial Y_a\over\partial s}
\epsilon_{ab}(-\half B\epsilon_{bc}Y_c)\cr &= -\pi \alpha n_B eB\int dt\, |{\bf
X}^B|^2,\cr}\eqn\Adphi$$
where we have specialized to the case of a uniform magnetic field in the
symmetric gauge ${\bf A} = \half{\bf B}\times{\bf r}$.    The integrals over
the boundaries at infinity vanish because the integrands fall off quickly
enough.  The integrals over the small circular boundaries $S_T$ have been
neglected as being small.  The diagonal self-energy terms in the  sums
\dphidphi\ do not depend on the collective coordinates and have been dropped.
Up to total time derivative terms,  the full collective vortex action is
$$\eqalign{ S_V &=\int dt\,\bigg(\sum_A \pi n_A\rho_0 \epsilon_{ab}X^A_a{\dot
X}^A_b\cr &\phantom{=}\,- \sum_{A < B} 2\pi\alpha n_An_B \ln|{\bf X}^A - {\bf
X}^B|^2\cr &\phantom{=}\,-\sum_A \pi\alpha n_A eB |{\bf X}^A|^2\biggr).\cr}
\eqn\voraction$$

\chapter{ Vortex Dynamics }

\section{ Classical Motion }
The classical equations of motion following from \voraction,
$$ \rho_0\dot X^A_a = 2\alpha\sum_{B\neq A}\,n_B\epsilon_{ab}
{X^B_b - X^A_b \over |{\bf X}^A - {\bf X}^B|^2}
- \alpha eB\epsilon_{ab}X^A_b,\eqn\classEoM$$
have the interesting property that they are first order in time and not
translation invariant.
There is one obvious constant of the motion for the equations of motion
\classEoM, which is essentially the angular momentum,
$$ {\sl d\over dt}\sum_A n_A|{\bf X}^A|^2 = 0. \eqn\const $$
Since the vortex Hamiltonian function,
$$H_V
= 2\pi\alpha \sum_{A< B} n_An_B \ln|{\bf X}^A - {\bf X}^B|^2
+\pi\alpha \sum_A n_A eB \,|{\bf X}^A|^2, \eqn\Hamiltonian $$
is a constant of the motion by virtue of the equations of motion
$$ 2\pi n_A\rho_0 {{\sl d} X_a^A\over {\sl dt}_{\phantom{a}}} = - \epsilon_{ab}
{\partial H_V\over\partial X^A_b},\eqn\canonical $$
there is one other constant of the motion, involving the separations of the
vortices,
$${\sl d\over dt}\prod_{A<B}|{\bf X}^A-{\bf X}^B|^{2n_An_B} = 0.\eqn\constII
$$

The motion of a single vortex in the absence of other vortices is a circular
orbit centered on the  origin, with angular velocity proportional to the
external magnetic field.  If a background of vortices of uniform density is
added, a vortex will move with an angular velocity which is the sum of two
pieces. One piece is proportional to the external field and is identical to the
angular velocity of a vortex in the absence of other vortices and the other is
proportional to the total vortex number  enclosed by the orbit.  This can be
seen from the equations of motion \classEoM, by averaging the contributions
of each vortex inside the orbit.  Since the `field' is inversely proportional
to the separation, in two dimensions we can smear the vorticity uniformly over
the enclosed area and ignore the contributions from outside the orbit, just as
one can for fields inversely proportional to the square of the separation
in three dimensions.

Thus the influence of other vortices is just the same as a background magnetic
field.  Two vortices of the same vorticity, $n_A$, in the absence of an
external field or other vortices, will orbit each other with an angular
velocity
which is inversely proportional to the square of their separation.  The
vortices behave as if they carry $n_A/e$ units of flux at their cores and a
charge-to-mass ratio $e^*/m^* = -\alpha e /\rho_0$.  The relative angular
momentum of two vortices can be calculated using the equation of motion
\classEoM\ and the definition of the momentum as the derivative of the
Lagrangian \voraction\ with respect to velocity, $P^A_a = \partial
L/\partial\dot X^A_a$. In the case of only two vortices with the same vortex
numbers,  $n_A = n_B = n$, there is a very simple relation:
$$\eqalign{J^{\rm rel} = \half\epsilon_{ab}
\xi_a\left({n\pi\rho_0^2\over\alpha eB}\dot\xi_b +
{4n^2\pi\rho_0\over eB}
\epsilon_{bc}{\xi_c\over|\pmb{{$\xi$}}|^2}\right)
&=\half\epsilon_{ab}\xi_a(\half m^*\dot\xi_b + e^*{\cal
A}_b(\pmb{{$\xi$}})),\cr
\epsilon_{ab}\partial_a{\cal A}_b({\bf x}) &= {4\pi n\over e}\delta^{(2)}
({\bf x}),\cr}\eqn\JrelCS$$
where $\pmb{{$\xi$}}={\bf X}^A -{\bf X}^B$ is the relative separation of the
two vortices $A$ and $B$. The term ${\cal A}_b$ in \JrelCS\ is an effective
vector potential, which is that of a point flux of $2n/e$ units.
This also strongly suggests  that the vortices carry attached point fluxes in
addition to their charge.  The factor of
two arises because there are two vortices and we must add the
contribution of each.
In fact, using the equations of motion \classEoM,
and the definition of momentum, we find the suggestive relation
$$\eqalign{P^A_a &= {2\pi n\rho_0^2\over eB\alpha}{\dot X}^A_a -
{2\pi n\rho_0\over B}{\cal A}_a({\bf X}^A) -
{2\pi n\rho_0\over B}A^{\rm ext}_a({\bf X}^A),\cr
&= m^*{\dot X}^A_a + e^*{\cal A}_a({\bf X}^A) + e^*A^{\rm ext}_a({\bf X}^A),\cr
{\cal A}_a({\bf X}^A) &\equiv \sum_{B\neq A} {2n_B\over e} \epsilon_{ab}
{X^B_b - X^A_b \over |{\bf X}^A - {\bf X}^B|^2}.\cr}\eqn\PVrel$$
Arovas has argued\Ref\Arovas{D.P.~Arovas, in {\sl Geometric Phases in Physics},
A.~Shapere and F.~Wilczek, eds., (World Scientific, Singapore, 1989) p.\ 288.}
that the extra factor of two in the statistical gauge field ${\cal A}_a$ is
necessary because each particle carries a flux as well as a charge.

The equations of motion \classEoM\ are classical equations giving definite time
evolution to the centers of the vortices. Classically, of course, the motion of
a charge in the background of an infinitesimally thin flux tube is the same as
the motion in a field-free region.  The fact that the `classical' motion of a
vortex is influenced by the other vortices and the existence of the point flux
in the relation \PVrel\ are the first hints that the motion described by the
action \voraction\ is quantal and not classical.

\section{ Quantum Motion in Symmetric Gauge }

We have said that the true meaning of  the dynamics of \voraction\ is quantal.
In order to describe the quantum mechanics of \voraction,  one must take into
account its unusual canonical structure.  Unlike the standard Lagrangian for a
massive point particle, the vortex Lagrangian is first order in time, which
means that there are constraints on its phase space. These constraints are
second-class according to the classification of Dirac. \Ref\Dirac{P.A.M.~Dirac,
{\it Lectures on Quantum Mechanics}, Yeshiva  University, (Academic Press, New
York, 1967)\semi A.J.~Hanson, T.~Regge and C.~Teitelboim, {\it Constrained
Hamiltonian Systems}, Accademia Nazionale dei Lincei (Rome, 1976).}  They are
$$ \varphi^A_a({\bf P}^A,{\bf X}^A) = P^A_a +  \pi n_A\rho_0\epsilon_{ab}X^A_b
\approx 0,\eqn\constraints $$
where $P^A_a $ is the momentum conjugate to the position of the \Ath vortex,
$X^A_a$.  The Poisson brackets of the constraints \constraints\ are
$$ \left\{\varphi^A_a, \varphi^B_b\right\} = 2\pi n_A\rho_0\delta^{AB}
\epsilon_{ab}.\eqn\Poisson $$
The Hamiltonian, which must conserve the constraints, is only defined  by its
numerical value on the phase space hypersurface defined by the constraints.
This leaves its algebraic expression in terms of the original unconstrained
phase space variables quite ambiguous.  Usually it is said that the Hamiltonian
is defined only up to the addition of the constraints with Lagrange
multipliers, $H^*_V = H_V^{\phantom{*}} + \lambda^A_a\varphi^A_a$.   If we take
this as the most general Hamiltonian for now, the conservation of the
constraints fixes the Lagrange multipliers, resulting in the Hamiltonian
$$\eqalign{ %
H_V^* &= 2\pi\alpha \sum_{A< B} n_An_B \ln|{\bf X}^A - {\bf X}^B|^2\cr
&\phantom{=}\, + {2\alpha\over\rho_0}\sum_{A\neq B}n_B\epsilon_{ab}P^A_a {X^B_b
- X^A_b \over|{\bf X}^A - {\bf X}^B|^2}\cr &\phantom{=}\, -
{eB\alpha\over\rho_0}\sum_A P^A_aX^A_b\epsilon_{ab}.\cr}\eqn\DiracHam$$
We have dropped an irrelevant constant term in the Hamiltonian \DiracHam, which
we could have included.

The constraints \constraints\ need to be taken into account in the quantum
mechanics in order to construct physical states for the system.   One possible
solution is to solve them by choosing, say, the $Y$ variables to  be the
momenta conjugate to the $X$ coordinates.  Then the states would be functions
of $X$ only.  We choose not to do this, but to implement the constraints as
operator conditions instead.   It is convenient to go to complex coordinates in
the case of the symmetric gauge.   We write
$$\eqalign{ z^A &= X_1^A + i X_2^A,\cr
 \overline{z}^A &= X_1^A - i X_2^A,\cr}\qquad
\eqalign{  -i\partial^A  &= \half\left(P^A_1 - iP^A_2\right),\cr
  -i\overline{\partial}^A  &= \half\left(P^A_1 + iP^A_2\right).\cr}\eqn\cplx
$$

For the motion of a single vortex in the background field we note that the
constraints, when written in complex form
$$\eqalign{\varphi &= \varphi_1 + i\varphi_2 = -2i\left(\overline{\partial}
+\half\pi\rho_0n z\right),\cr \overline\varphi &= \varphi_1 - i\varphi_2 =
-2i\left({\partial} -\half\pi\rho_0n \overline{z}\right),\cr}\eqn\cplxconst $$
have Poisson brackets
$$\left\{\varphi, \overline\varphi\right\} = -4\pi i n \rho_0.\eqn\PoissonII$$
These Poisson brackets become the commutator relations of creation and
annihilation operators upon quantization.  Thus, we may impose the constraints
by asking that the correct one of them annihilate physical states.  In the case
of a single vortex of vorticity $n = 1$, we impose $\varphi$ and take the
states  to be eigenstates of the Hamiltonian \DiracHam.
$$\eqalign{ \hat\varphi\,\Psi(z,\zbar) &= -2i\left(\overline{\partial}
+\half\pi\rho_0 z\right)\Psi(z,\zbar) = 0,\cr \hat H^*\Psi(z,\zbar) &=
{eB\alpha\over \rho_0}\left(z\partial -
\zbar\overline{\partial}\right)\Psi(z,\zbar)=E\Psi(z,\zbar).\cr}\eqn\states$$
The resulting one-vortex physical states are
$$\eqalign{\Psi(z,\zbar) &= z^N\exp( -\half\pi\rho_0|z|^2 ),\cr N &= {\rho_0
E\over eB\alpha}.\cr}\eqn\eigenstate $$
Being purely analytic functions times an exponential of $|z|^2$, these states
are of the form of the lowest Landau level.  Because they are also eigenstates
of an annihilation operator,  these states are of the form of a coherent
excitation in the $N^{\rm th}$ Landau level as well.  One needs more evidence
to decide which is the proper interpretation.    Others have found that motion
restricted to the lowest Landau level can be related to one-dimensional
fermions through an action that is first order in time,\Ref\Iso{S. Iso, D.
Karabali and B. Sakita, {\it One-dimensional Fermions as Two-dimensional
Droplets via Chern-Simons Theory\/}, preprint CCNY-HEP-92/1,
hepth@xxx/9202012.}
and we will also come to the conclusion that the lowest Landau level
interpretation is most natural after considering the motion of vortices in a
half plane.

We note here that positive and negative winding number vortices behave as if
they have opposite charge in this quantization, in that their wave functions
are complex conjugates of each other.  This is in contrast to the classical
motion, where one cannot tell the difference between positive and negative
winding number vortices by their direction of motion because all vortices orbit
the center in the same sense.

\section{ The Motion of Two Vortices }

The equations of motion \classEoM\ determine that two vortices of the same
vorticity $n$ will orbit each other at constant separation $|\pmb{{$\xi$}}|^2
= |{\bf X}^{(1)} - {\bf X}^{(2)}|^2$ with constant relative angular velocity
$$\dot\theta = {2\alpha n\over \rho_0|\pmb{{$\xi$}}|^2} +
{\alpha\over\rho_0}eB.\eqn\AVrel
$$
The center of mass, ${\bf X}_{\rm CM} = \half({\bf X}^1 + {\bf X}^2)$, will
orbit the origin at constant radius and angular velocity
$$\dot\Theta ={\alpha\over\rho_0}eB.\eqn\AVcm$$
The Hamiltonian separates in CM and relative coordinates,
$$ H = H_{\rm CM} + H_{\rm rel} = 2\pi\alpha eB\,|{\bf X}_{\rm CM}|^2 +
2\pi\alpha\ln|\pmb{{$\xi$}}|^2 + \half\pi\alpha eB|\pmb{{$\xi$}}|^2,\eqn\Hsep
$$
as do the constraints
$$\eqalign{ \varphi^{\rm CM}_a &= P^{\rm CM}_a + 2\pi\rho_0 n\epsilon_{ab}
X^{\rm CM}_b\approx 0,\cr
\varphi^{\rm rel}_a &= P_{\xi\,a} + \half\pi\rho_0 n\epsilon_{ab}
\xi_b\approx 0.\cr }\eqn\CONsep$$
If the usual complex combinations of constraints are made
$$\eqalign{\varphi^{\rm CM} &= \varphi^{\rm CM}_1 + i\varphi^{\rm CM}_2
\approx 0,\cr
\varphi^{\rm rel} &=\varphi^{\rm rel}_1 + i\varphi^{\rm
rel}_2\approx 0,\cr}\eqn\cplxsep$$
and imposed on states, one finds that the states are all of the form
$$\Psi(\xi,Z) = \xi^\theta Z^N\exp\left( -\pi\rho_0n|Z|^2
- {\pi\rho_0 n\over 4}|\xi|^2 \right),\eqn\twovorstate$$
where $Z = X^{\rm CM}_1 + i X^{\rm CM}_2$, $\xi = \xi_1 + i \xi_2$, and
$N$ must be an integer if the center of mass wave function is to be
single-valued.  Otherwise, the values of $\theta$ and $N$ are completely
undetermined by the Hamiltonians
$$ \eqalign{ H^*_{\rm CM} &\sim Z\partial_Z
- \overline{Z} \partial_{\overline{Z}} ,\cr
H^*_{\rm rel} &\sim \ln(\xi\partial_\xi -
\overline{\xi}\partial_{\overline{\xi}})
        +(\xi\partial_\xi - \overline{\xi}\partial_{\overline{\xi}}).\cr}
\eqn\Hops$$

\REF\Chiao{R.Y. Chiao, A. Hansen and A.A. Moulthrop, \prl {\bf 54}
(1985) 1339\semi A. Hansen, A.A. Moulthrop and R.Y. Chiao, \prl {\bf 55} (1985)
1431.}

Thus, from the point of view of this quantization, any statistics are allowed.
This result is less restrictive than those obtained in Ref.~[\Chiao].

\section{ Boundary Conditions and Gauge Invariance }

We have not yet touched upon boundary conditions for the Ginzburg-Landau
equations \psimotion.  Because the order parameter is essentially a macroscopic
wave function, the boundary conditions should ensure that no current flows
across free boundaries of a sample.  This is guaranteed by the usual
Ginzburg-Landau boundary conditions,
$$ \hat{\bf n}\cdot{\partial{\cal H}_{GL}\over\partial(\grad\overline\psi)}
\bigg|_{\partial S}= 0,\eqn\BCi$$
which follow from considering the minimization of the free energy with respect
to arbitrary variations $\delta\psi$ which do not vanish on the
boundaries of the sample. This boundary condition can be thought of as
determining the correct gauge for a particular sample geometry.  In an infinite
sample, the multi-vortex ansatz was that the vortices were superimposed upon a
stationary background quantum fluid `sea.'  In the case of a finite system, the
correct ansatz is
$$\phi=\sum_An_A\psi_A({\bf x},{\bf X}^A)+\omega({\bf{x}}),\eqn\multivorII$$
so that in the absence of vortices, the boundary conditions on the quantum sea
$\omega$ will be
$${\bf n}\cdot\left(\grad\omega - e{\bf A}_{\rm ext}\right)\big|_{\partial S} =
0.\eqn\BCii$$
This is equivalent to using $\omega$ to change the gauge of ${\bf A}_{\rm ext}$
so that it will satisfy
$$\eqalign{ \grad\cdot{\bf A}_{\rm ext} &= 0,\cr \hat{\bf n}\cdot{\bf A}_{\rm
ext}\big|_{\partial S} &= 0,\cr} \eqn\BCgauge$$
and leaving the multi-vortex ansatz to satisfy the Laplace equation with
Neumann boundary conditions
$$\eqalign{\grad^2\psi_A({\bf x}) &= 0,\cr \hat{\bf n}\cdot\grad\psi_A
\big|_{\partial S} &= 0.\cr}\eqn\multivorBC$$
In simple geometries one can solve the boundary condition in \multivorBC\ by
the method of images,
$$\psi_A({\bf x}, {\bf X}^A) = \phi_A({\bf x}, {\bf X}^A) +
\sum_j \nu_j\phi_A({\bf x}, {\bf R}_j({\bf X}^A)).\eqn\Image$$
Here $\nu_j$ is the image vortex strength, whose sign depends on the number
of reflections the original vortex at ${\bf X}^A$ has undergone to produce
the image vortex at ${\bf R}_j({\bf X}^A)$.

In the derivation of the action \voraction, the symmetric gauge was chosen.
This choice is forced if one imagines the infinite sample to be the limit of
a circular sample when the radius increases without bound.  The vector
potential must be in the symmetric gauge and vanish at the
center of the circle.

As a simple example, we examine the motion of vortices in a constant magnetic
field and in a finite circular sample.  We then specialize to the case
of a single vortex.  The derivation of the action starts from the multi-vortex
ansatz
$$\eqalign{\phi({\bf x},t) &= \sum^N_{A=1} n_A\left[
\phi_A({\bf x}, {\bf X}^A) - \phi_A({\bf x}, {\bf R}({\bf X}^A))\right],\cr
{\bf R}({\bf X}^A) &= {r^2\over |{\bf X}^A|^2}{\bf X}^A,\cr}\eqn\multivorIII$$
where $r$ is the radius of the disk and the center of the disk is the center of
the coordinate system.  We proceed as before, substituting the ansatz
back into the action and putting cuts from each vortex to a {\sl fixed} point
on the outer boundary of the sample.  The resulting action is
$$\eqalign{S_V&=\int dt\,\bigg(\sum_A\pi n_A\rho_0\epsilon_{ab}X^A_a
{\dot X}^A_b\cr
&\phantom{=}\,- \sum_{A < B} 2\pi\alpha n_An_B \ln|{\bf X}^A -
{\bf X}^B|^2\cr
&\phantom{=}\,+ \sum_{A}\sum_B\pi\alpha n_An_B \ln|{\bf X}^A -
{\bf R}({\bf X}^B)|^2\cr
&\phantom{=}\,-\sum_A\pi\alpha
n_AeB |{\bf X}^A|^2\biggr).\cr}\eqn\voractionII$$

The classical equation of motion for a single vortex of vorticity $n$, is
$$ \dot X_a = -\epsilon_{ab}X_b{\alpha\over\rho_0}\left(eB + {n\over r^2 -
|{\bf X}|^2}\right).\eqn\bdryEoM$$
The quantum states are again found to be of the same form as \eigenstate,
$$\Psi(z,\zbar) = z^N\exp( -\half\pi\rho_0|z|^2 ),\eqn\eigenII$$
but now the dependence of the energy on $N$ is considerably more complicated.

\section{ Quantum Motion in a Half Plane }

In choosing the boundary conditions for motion in a half plane, we must be
careful to note that there is not a unique vector potential ${\bf A}^{\rm ext}$
satisfying the conditions \BCgauge.    We take our geometry to be the upper
half plane $y>0$.  It is most natural to take the Landau gauge condition $A_x =
-By$, $A_y = 0$.  The non-uniqueness of the vector potentials comes in through
the addition of a constant vector field $A_x = C$, $A_y = 0$.  If the limit is
taken by starting with a rectangle and letting three of its sides go to
infinity, then the constant vector potential is infinite, unfortunately.  It is
still useful to look at the example to determine the behavior of vortices near
a long straight boundary far from other boundaries.

Each vortex in this geometry has a single reflected image,
${\bf{R}}(X,Y)=(X,-Y)$.  Constructing the vortex action as in the circular
case, we find
$$\eqalign{S_V&=\int dt\,\bigg(\sum_A\pi n_A\rho_0\epsilon_{ab}X^A_a {\dot
X}^A_b\cr &\phantom{=}\,-\sum_{A<B}2\pi\alpha n_An_B \ln|{\bf X}^A-{\bf
X}^B|^2\cr  &\phantom{=}\,+ \sum_{A}\sum_B\pi\alpha n_An_B \ln|{\bf X}^A -
{\bf R}({\bf X}^B)|^2\cr  &\phantom{=}\,-\sum_A 2\pi\alpha n_A eB
(X_2^A)^2\biggr).\cr} \eqn\voractionIII$$%
The action has the same constraints as before, but the Hamiltonian is
different.  The single-vortex Hamiltonian,
$$ H_V = -\pi\alpha n^2\ln|{\bf X} -  {\bf R}({\bf X})|^2+ 2\pi\alpha neB
(X_2)^2,\eqn\oneVortHam$$
must be modified to preserve the constraints.  We use the ``star variables''
explained in appendix A. For vortex number $n=1$,
$$\eqalign{H_V^*&=H_V(X_2^*)=2\pi\alpha eB(X_2^*)^2-\pi\alpha
\ln[4(X_2^*)^2],\cr  X_2^* &= \half(X_2 -
{1\over\pi\rho_0}P_1).\cr}\eqn\starHam$$
The star variable $X_2^*$ commutes with both constraints so the Hamiltonian
does also.  Diagonalizing  $X_2^*$ will diagonalize $H_V^*$ as well, so we
solve the simultaneous conditions
$$\eqalign{ \hat\varphi\Psi &= 0,\cr \hat X_2^*\Psi &=
k\Psi,\cr}\eqn\nsstate$$
to find the physical states which diagonalize the Hamiltonian \starHam,
$$ \Psi({\bf X}) = e^{i\pi\rho_0X_1X_2}\exp\left({-i\pi\rho_0
kX_1-{\pi\rho_0}(X_2-{k\over 2})^2}\right).\eqn\nsLandau$$
Except for the extra factor $e^{i\pi\rho_0X_1X_2}$, $\Psi$ is a Landau gauge
ground state.  Being a pure phase, this extra factor is physically irrelevant.

\chapter{ Magnetic Translations and Theta Terms }

A interesting fact of the Landau problem is that while a uniform field strength
is translation invariant, one must choose a gauge potential in order to  write
the usual minimally coupled Hamiltonian
$$ H = {1\over 2m}\left({\bf p} - e{\bf A}\right)^2,\eqn\LHam$$
thereby breaking manifest translation invariance.  The translation invariance
is not really lost, however.  A combination of a translation
$$ {\bf x}\longmapsto{\bf x}^{\prime} = {\bf x} + {\bf a}\eqn\trans$$
and a gauge transformation,
$${\bf A}=\half{\bf B}\times{\bf x}\longmapsto{\bf A}^{\prime}= {\bf
A}+\grad(\half{\bf x}\cdot{\bf B}\times{\bf a}),\eqn\gaugetrans$$
will leave the Hamiltonian invariant.  The combination of transformations
\trans\ and \gaugetrans\ together are a magnetic translation.  Another way of
getting at the translation invariance is to look at the action
$$S=\int dt[\half m(\dot{\bf x})^2+e{\bf A}\cdot\dot{\bf x}],\eqn\Lagrangian$$
in a linear gauge ${\bf A}({\bf x}) = {\ss K}\cdot{\bf x}$.  Under a
translation ${\bf x} \mapsto {\bf x} + \delta{\bf x}$ the action changes by
$$\eqalign{\delta S &= \int dt {\sl d\over dt}\left(\delta{\bf x}\cdot{\ss
K}\cdot{\bf x}\right)\cr &= \int dt\left(\delta\dot{\bf x}\cdot{\partial
L\over\partial\dot{\bf x}} +  \delta{\bf x}\cdot{\partial L\over\partial{\bf
x}}\right)\cr &= \int dt{\sl d\over dt}\left(\delta{\bf x}\cdot{\partial L\over
\partial\dot{\bf x}}\right).\cr}\eqn\Schange$$
The first equality is by direct computation, the second is the chain rule,
 and the third uses the equations of
motion. This implies that there is a conserved quantity for the system
$$0={\sl d\over dt}\left(\delta{\bf x}\cdot\left[{\partial
L\over\partial\dot{\bf x}} - {\ss K}\cdot{\bf x}\right]\right)  = {\sl d{\it
Q}\over dt}.\eqn\MagTranQ$$

The action \voraction\ is not quite invariant even under magnetic translations,
however.  The classical orbits have their centers shifted by  magnetic
translations, but under a magnetic translation by the vector $\bf a$, the
vortex action transforms as
$$\eqalign{S^{\prime}_V(\{{\bf X}^{\prime A}-{\bf a}\})&=S_V(\{
{\bf X}^{\prime A}\}) + \Gamma[{\bf a}, \{{\bf X}^{\prime A}\}],\cr
\Gamma[{\bf a}, \{{\bf X}^{\prime A}\}]&=-\sum_{A}n_A\pi\rho_0\int
dt\,{\sl d\over dt}\left(\epsilon_{ab}{\rm a}_aX^{\prime A}_b\right),\cr}
\eqn\gaugetheta $$
where $S^{\prime}_V$ denotes the gauge transformed action.  A similar
analysis to \Schange\ and \MagTranQ\ performed for the action \voraction,
shows that there is no conserved charge even under a {\sl magnetic} translation
of all vortex centers simultaneously:
$$Q=\left({\bf a}\cdot\sum_A\left[{\partial L\over\partial\dot{\bf X}^A}
- \pi\rho_0\varepsilon\cdot{\bf X}^A \right]\right) \equiv 0.\eqn\MagTrQii$$
Here $\varepsilon$ is the matrix $\epsilon_{ab}$.

If a theta term,  $\Gamma[{\bf a}, \{{\bf X}^{A}\}]$, as in \gaugetheta, is
present in the action, then the constraints are modified, as is the Hamiltonian
$H^*_V$:
$$\eqalign{\hat\varphi &= \overline{\partial} +  \half\pi\rho_0(z-z_0)\approx
0,\cr \hat{\bar{\varphi}} &= \partial -\half\pi\rho_0(\zbar-\zbar_0)\approx
0,\cr \hat{H}^* &= z\partial - \zbar\overline{\partial} +
\half\pi\rho_0(z\zbar_0 + \zbar z_0).\cr}\eqn\ThetaOps$$
Here we have taken $z_0 = {\rm a}_1 + i{\rm a}_2$.  The general physical state
for a single vortex of winding number $n=1$ is then of the form
$$\Psi(z,\zbar) = C z^N \exp\left(-\half\pi\rho_0(|z|^2 + z\zbar_0 -\zbar z_0)
\right).\eqn\thetastate$$

The shifted states do not stay in a single Landau level but are mixtures of all
Landau levels. The only effect of magnetic translations is to shift the theta
terms, which leaves the classical equations unaffected, but the quantum states
are radically changed.  Even  when no field is present and magnetic
translations become ordinary translations, there is no symmetry of the theory,
so it is not surprising that a magnetic translation of \voraction\ should mix
Landau levels.

\chapter{ Discussion and Conclusions }

We have demonstrated, under the assumptions of an appropriate Landau free
energy, that vortex defects in the order parameter are charged  and that these
vortices obey an equation of motion which is first order in time. After
quantization these vortices are seen to behave correctly quantum mechanically,
and, in fact, can be taken to occupy the lowest Landau level of the minimally
coupled single-vortex quadratic Hamiltonian
$$H={1\over 2m^*}\left({\bf p}-e^*{\bf A}_{\rm ext}\right)^2,\eqn\quadHam$$
with $e^*/m^* = -\alpha e/\rho_0$ and $e^* = 2\pi\rho_0/B$.  We have not
postulated a form for the  potential $V(|\psi|^2)$ in our Landau free energy
\genH, but have assumed that  it can be chosen so that the minimum will set the
parameter $\rho$ to  $\rho_0 = {eB\over 2\pi (2n + 1)}$. The constant $\alpha$
can be chosen to fix the effective mass $m^*$.

All the vortices behave classically, according to \classEoM, as if they  have
the opposite sign charge to the background quantum fluid.  That is, they all
travel in the same direction regardless of their vorticities $n_A$.  The
collective Hamiltonian function for the vortices \Hamiltonian, has a direct
connection to the Laughlin many-body wave function.  The relation between
\Hamiltonian\ and the Laughlin wave function is
$$ |\Psi_{\rm Laughlin}(\{Z_A\})|^2 \propto \exp\left( -{1\over2\pi\alpha m}
H_V(\{Z_A\})\right).\eqn\LaughlinRel$$
When all the vortex numbers are taken to be positive, $n_A > 0$, some $n_A = m$
and some $n_B = 1$, and the magnetic field is in the $-{\bf \hat z}$ direction,
\LaughlinRel\ is the modulus of the Laughlin wave function of electrons and
quasiholes of charge $1/m$.

Numerical work, solving the first order equations of motion, \classEoM, for a
random initial configuration of particles, all with the same vortex numbers,
shows that the system evolves so that the vortices form a lattice of nearly
uniform density which rotates rigidly in the effective magnetic field which is
the background field plus  that produced by the vortices
themselves.\Ref\Fosdal{ S. Fosdal, private communication.}  Thus it seems
reasonable to write an effective Ginzburg-Landau theory for the fluid of
vortices, in which further vortices may form.  This lends support to the
correctness of the phenomenological theory.  In this way one may hope to
reproduce the whole hierarchy of fractional quantum Hall states through the use
of a Ginzburg-Landau theory.

The present work assumes the existence of multi-vortex solutions which have
core sizes much smaller than their separations.  We have assumed that such
solutions exist for the appropriate choice of functions $f_i(|\psi|^2)$, as
they are known to exist in the Abelian Higgs model.\Ref\Taubes{C.H.~Taubes,
{\it Commun. Math. Phys.} {\bf 72} (1980) 277.}

We have not fixed all the parameters in the theory to relate the average
density of vortices to the magnetic flux through the sample.  It seems
plausible that this could be done.  Instead of choosing finite energy as our
criterion, we have chosen that the equations of motion for the non-linear
Schr\"odinger equation have constant density solutions.  Because of the
unscreened external vector potential,  it is impossible to keep just the first
term in \genH\ and demand finite energy in an infinite sample.  It is our hope
that including more of the terms in the free energy \genH\ will yield a
detailed macroscopic description of the fractional quantum Hall effect,
determining, for instance, the vortex density from the applied magnetic field,
without harming the basic picture of vortex dynamics we have sketched.

{}From the point of view of first order vortex dynamics, it is unclear how the
statistics of the quantum vortices are determined.   From the discussion
leading up to eq.\ \twovorstate, it seems as though the quantum statistics can
be chosen  independently of the effective charge, but there are tantalizing
hints, from eq.\ \JrelCS\ or from \PVrel, that the Chern-Simons induced
statistics directly related to the effective charge are already implied.

\ack

It is a pleasure to thank Dennis Crossley for a useful discussion,
A.P.~Balachandran, B.~Durand, and L.~Durand for reading the manuscript and
making useful suggestions on its improvement and Steve Fosdal for sharing with
us his numerical simulations of many  vortex motion. This work was supported in
part by DOE grant No.\ DE-AC02-76-ER00881.

\Appendix{A}
\centerline{\caps Dirac Quantization of Second-Class Systems }

When an action has second-class constraints, $\varphi_\alpha \approx 0$, there
are several options for constructing a quantum system from the classical
system.  If the constraints are simple enough, one can just solve them and
write the dynamics in terms of a reduced set of variables.  If one chooses not
to do this, there are still several possibilities.  Since the Poisson bracket
of the constraints does not vanish,
$$\{\varphi_\alpha, \varphi_\beta\} = \Delta_{\alpha\beta},
\quad\det\Delta\neq 0, \eqn\SCPoisson$$
the constraints cannot simply annihilate the quantum states, or the theory will
be trivial,
$$\widehat{\varphi}_\alpha\ket{\psi}=0\quad\Rightarrow\quad\ket{\psi}=0.
\eqn\inconsis$$
One can, however, require that the constraints have vanishing matrix elements
between any two physical states,
$$\VEV{\psi^{\prime}|\widehat{\varphi}_\alpha|\psi} = 0.\eqn\GuptaBleuler $$

A stricter condition is to require that some complex combinations of the
constraints annihilate all physical states,
$$\eqalign{C^A_\alpha\widehat{\varphi}_\alpha\ket{\psi} &= 0,\cr
C^A_\alpha\Delta_{\alpha\beta}C^B_\beta &= 0,\cr
\det_{(AB)}[C^A_\alpha\Delta_{\alpha\beta}\overline{C}^B_\beta] &\neq 0.\cr}
\eqn\PseudoDirac $$
In this case, one must make sure that the Hamiltonian preserves the constraints
$C^A_\alpha\varphi_\alpha$:
$$\{C^A_\alpha\varphi_\alpha, H\} = V^A_BC^B_\alpha\varphi_\alpha.\eqn\CH$$

One possibility is to use so-called ``star variables'' when the matrix
$\Delta_{\alpha\beta}$ is constant.  The idea is to transform  from the old set
of canonical variables $z_a$ to a new set of variables $z^*_a$ (here the star
means modified variables, not complex conjugated variables)
$$\eqalign{z_a^* &= z_a - \varphi_\alpha(\Delta^{-1})^{\alpha\beta}\{
\varphi_\beta,z_a\},\cr &\{z^*_a,\varphi_\alpha\}\approx
0.\cr}\eqn\starvariables $$
Any function can be made to have weakly vanishing Poisson bracket with the
constraints simply by replacing the function of the old canonical variables by
the same function of the star variables,
$$\{F(z^*), \varphi_\alpha\} \approx 0. \eqn\Invariant$$
The weak equalities in \Invariant\ and \starvariables\ will be equalities when
the Poisson brackets $\{\varphi_\beta,z_a\}$ as well as $\Delta_{\alpha\beta}$
are constants.

\Appendix{B}
\centerline{\caps Spectrum and States of Landau Problem }

The electron states in a two dimensional conductor with a perpendicular
magnetic field are highly degenerate.  These degenerate states, the Landau
levels, can easily be found in the Landau gauge
$$ {\bf A}(x,y) = -By \,{\bf \hat x}.\eqn\nonsymgauge $$
The momentum $p_x$ is a conserved quantity of the Landau Hamiltonian
$$ H = {1\over 2m}({\bf p} - e{\bf A})^2 = {1\over 2m}[ (-i\partial_x + eBy)^2
+ (-i\partial_y)^2 ],\eqn\nonsymmagham$$
as is the quantity $p_y + eBx$.  These are the magnetic translation
operators.  By diagonalizing $p_x$, one finds that the
energy eigenstates are products exponentials and harmonic oscillator wave
functions, $\phi_n$, centered at $y_0 = -{k\over eB_0}$ with frequency
$eB_0\over m$;
$$\psi_n = e^{ikx}\phi_n(y - y_0).\eqn\nsymlanlev$$

In the symmetric gauge
$$ {\bf A}({\bf r}) = \half {\bf B}\times{\bf r},\eqn\symgauge$$
the Hamiltonian is
$$ \eqalign{ H &= {1\over2m}\{\Pi_x^2 + \Pi_y^2\} ={1\over 4m}\{ \Pi_z\Pi_{\bar
z} + \Pi_{\bar z}\Pi_z\}\cr &= {1\over 2m}\bigl\{-4\partial\parbar
+({eB_0\over 2})^2 |z|^2 - eB_0(z\partial -
\zbar\parbar)\bigr\},\cr}\eqn\symmagham$$
where ${\bf\Pi} = -i\nabla - e{\bf A}$.  The problem has been formulated in
complex coordinates as in eq.\ \cplx. The $\Pi_z$ and $\Pi_{\zbar}$ can be
treated as creation and annihilation operators.  They satisfy the commutation
relations
$$ [\Pi_z,\Pi_\zbar] = 2m\omega.\eqn\Picom$$
Here $\omega$ denotes the signed frequency $eB_0\over m$. The sign of $\omega$
determines which of $\Pi_z$ and $\Pi_\zbar$ is the creation operator.  In the
following $\omega > 0$.  The notation is simpler if the complex coordinates are
scaled by $(m|\omega|)^{-1/2}$. The spectrum generating operators can then be
identified as
$$\eqalign{ \hat a^\dagger  &= -{i\over \sqrt{2m|\omega|}} \Pi_\zbar =
\sqrt{2}(-\partial + \textstyle{1\over 4}\zbar),\cr \hat a&= {i\over
\sqrt{2m|\omega|}}\Pi_z = \sqrt{2} (\parbar + \textstyle{1\over
4}z).\cr}\eqn\specgenops$$
Just as for the one dimensional harmonic oscillator, the ground state is
annihilated by $\hat a$, otherwise it could be lowered to a state of lower
energy and the Hamiltonian must be bounded below. It follows that the ground
states of the Hamiltonian are of the form
$$\eqalign{\psi_f(z, \zbar) &= \exp(-{|z|^2/4}) f(z),\cr H\psi_f &=
\half{|\omega|}\psi_f.\cr}\eqn\gndstate$$
Excited states are built upon any ground state \gndstate, by applying creation
operators.
$$\ket{n,f} = {1\over\sqrt {n!}}(\widehat{a}^\dagger)^n \exp(-{|z|^2 / 4})
f(z).\eqn\excitedstates$$
There are again two constants of the motion for this problem.  These are the
magnetic translation operators
$$\eqalign{\widehat{b} &=\sqrt{2}(\partial + \textstyle{1\over 4}\zbar),\cr
\widehat{b}^{\dagger} &=\sqrt{2}(-\parbar + \textstyle{1\over4}z),\cr}
\eqn\magTranOps$$
which commute with the spectrum generating operators,
$$[\widehat{a}, \widehat{b}] = [\widehat{a}, \widehat{b}^\dagger]
= [\widehat{a}^\dagger, \widehat{b}] =
[\widehat{a}^\dagger, \widehat{b}^\dagger] = 0.\eqn\commuteAB$$
The full set of states can be generated from a {\sl single} ground state
through the use of the two raising operators
$$\eqalign{\ket{n,m}&={1\over\sqrt{n!m!}}(\widehat{a}^\dagger)^n
(\widehat{b}^\dagger)^m\exp(-{|z|^2 / 4}),\cr
\widehat{H}\ket{n,m} &= (n + \half)|\omega|\ket{n,m}.\cr}\eqn\AllLandauStates$$

\refout
\vfill
\eject
\leftline{\fourteenbf Figure Caption }

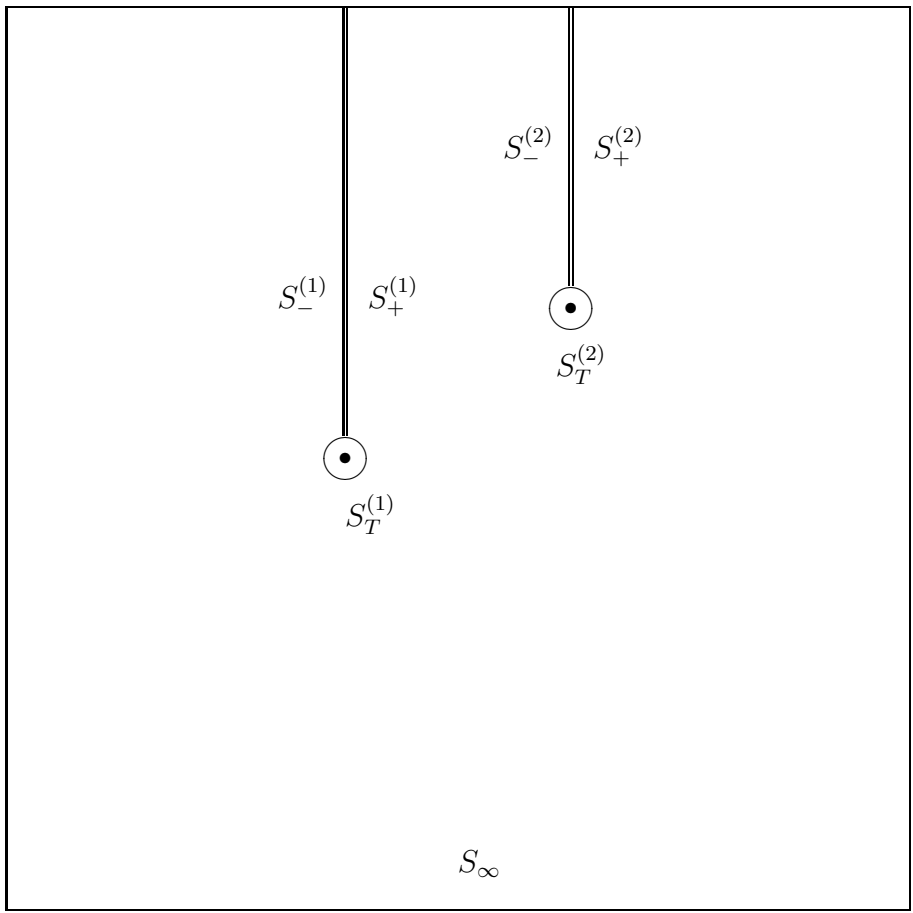
Figure 1: The integration region and boundary cuts
for a two vortex configuration.
\vfill
\bye